*Article*

# Why does tinnitus vary with naps?

## *A polysomnographic prospective study exploring the somatosensory hypothesis.*

**Robin Guillard [1,2]\*, Vincent Philippe [2], Adam Hessas [2], Brice Faraut [3-4], Sarah Michiels [5], Minchul Park [6], Marco Congedo [7], Alain Londero [8] and Damien Léger [3-4]**

[1]  GIPSA-Lab, Univ. Grenoble Alpes, CNRS, Grenoble INP, Grenoble, France.
[2]  Robin Guillard EIRL, Grenoble, France.
[3]  Université Paris Cité, VIFASOM ERC 7330, Vigilance Fatigue Sommeil et Santé publique, Paris, France.
[4]  APHP Hôtel-Dieu, Centre du Sommeil et de la Vigilance, Paris, France
[5]  REVAL Rehabilitation Research Center, Faculty of Rehabilitation Sciences, Hasselt University, Hasselt, BE Belgium
[6]  MinChul Park, ORCID: 0000-0001-5500-1623, University of Canterbury, Christchurch, New Zealand
[7]  Grenoble Alpes University, CNRS, Grenoble INP
[8]  Université Paris Cité, Institut Pasteur, AP-HP, Hôpital Lariboisière, Service ORL, Unité Explorations Fonctionnelles, INSERM, Fondation Pour l'Audition, IHU reConnect, F-75010 Paris, France ;

Correspondence: robin.guillard@grenoble-inp.fr, 17 boulevard de Picpus, 75012, Paris

*Article*

# Why does tinnitus vary with naps?
## *A polysomnographic prospective study exploring the somatosensory hypothesis.*

**Robin Guillard [1,2*], Vincent Philippe [2], Adam Hessas [2], Brice Faraut [3-4], Sarah Michiels [5], Minchul Park [6], Marco Congedo [1], Alain Londero [7] and Damien Léger [3-4]**

1   GIPSA-Lab , Univ. Grenoble Alpes, CNRS, Grenoble INP, Grenoble, France.
2   Robin Guillard EIRL, Grenoble, France.
3   Université Paris Cité, VIFASOM ERC 7330, Vigilance Fatigue Sommeil et Santé publique, Paris, France.
4   APHP Hôtel-Dieu, Centre du Sommeil et de la Vigilance, Paris, France
5   REVAL Rehabilitation Research Center, Faculty of Rehabilitation Sciences, Hasselt University, Hasselt, BE Belgium
6   MinChul Park, ORCID: 0000-0001-5500-1623, University of Canterbury, Christchurch, New Zealand
7   Grenoble Alpes University, CNRS, Grenoble INP
8   Université Paris Cité, Institut Pasteur, AP-HP, Hôpital Lariboisière, Service ORL, Unité Explorations Fonctionnelles, INSERM, Fondation Pour l'Audition, IHU reConnect, F-75010 Paris, France

Correspondence: robin.guillard@grenoble-inp.fr, 17 boulevard de Picpus, 75012, Paris

**Abstract:** (1) Background: Tinnitus, defined as the conscious awareness of a noise without any identifiable corresponding external acoustic source, can be modulated by various factors. Among these factors, tinnitus patients commonly report drastic increases of tinnitus loudness following nap sleep. Previous studies have suggested that this clinical pattern could be attributed to a somatosensory modulation of tinnitus. To our knowledge, no polysomnographic study has been carried out to assess this hypothesis.

(2) Methods: For this observational prospective study, 37 participants reporting frequent increases of tinnitus following naps were recruited. They participated to six full-polysomnography nap attempts over two days. Audiological and kinesiologic tests were conducted before and after each nap attempt.

(3) Results: 197 naps were collected. Each nap at each time of day elicited an overall significant increase in tinnitus minimum masking level (MML). Each inter nap period elicited an overall significant decrease. Tinnitus modulations were found significantly correlated with nap sleep duration (Visual numeric scale on tinnitus loudness, VNS-L, p< 0.05), with snoring duration (MML, p< 0.001), with snoring average sound level (VNS on tinnitus intrusiveness, VNS-I, p< 0.05) and with sleep apnea count (VNS-I, p< 0.001).

(4) Conclusions: This study confirms objectively that tinnitus may increase following naps. No association was found between these modulations and somatosensory modulations involving the temporomandibular joint and cervical areas. However, it may be possible that nap-induced tinnitus modulations are a hidden form of somatosensory modulation as snoring and sleep apnea events are often related to tensor veli palatini muscle dysfunction.

**Highlights:**

- Demonstration of tinnitus nap-induced modulations under polysomnographic control.

- Nap sleep duration, snoring and sleep apnea appear to correlate such modulations.

- Methodological robustness: patients are compared to themselves over 6 naps.

**List of abbreviations**

CRF : Case Report Form
ET : Eustachian Tube
ECG : ElectroCardioGraphy
EEG : ElectroEncephaloGraphy
EMG : ElectroMyoGraphy
EOG : ElectroOculoloGraphy
MML : Minimum Masking Level
NITM : Nap-Induced Tinnitus Modulation
SIT : Sleep-Intermittent Tinnitus
SPO2 : Blood-oxygen saturation level
THI : Tinnitus Handicap Inventory
TIAN : Tinnitus Increases After Naps
TMA : Tinnitus Momentary Assessment
TMM : TuboManoMetry
TT : Tensor Tympani
TVP : Tensor Veli Palatini
VNS-L : Visual Numeric Scale of tinnitus Loudness
VNS-I : Visual Numeric Scale of tinnitus Intrusiveness

## 1. Introduction

Tinnitus is defined as "the conscious awareness of a tonal or composite noise for which there is no identifiable corresponding external acoustic source" (De Ridder et al., 2021). This chronic symptom is widely prevalent, affecting approximately 14% of the world population, 2% of which in a severe form (Jarach et al., 2022).

Patients often report that perception of tinnitus can vary over time (Dauman et al., 2015; Probst et al., 2017). Multiple sources of tinnitus modulations have been identified: somatosensory modulations (Ralli et al., 2017, 2016; Shore et al., 2007), stress modulations (Elarbed et al., 2021), noise modulations (Roberts, 2007; Vernon and Schleuning, 1978), specific food intake modulations (Marcrum et al., 2022), atmospheric pressure modulations (Colagrosso et al., 2019)…

Another cause of modulation of tinnitus reported by roughly a third of the tinnitus population is sleep, and more specifically napping sleep (Levine and Oron, 2015; Ralli et al., 2016). Surprisingly, twice as many patients report increases of tinnitus following naps compared to decreases of tinnitus after naps (Guillard et al., 2023c; Guillard et al., 2024; van den Berge et al., 2017). Three studies have suggested from clinical observations that this pattern could be associated to somatosensory modulations of tinnitus (Levine, 2013; Levine and Oron, 2015; Ralli et al., 2016), and even that these modulations could be more frequent when tinnitus patients sleep in a seated position. In a recent cross-sectional database study aiming at understanding the symptomatological profile of tinnitus patients reporting these nap-induced tinnitus modulations (NITM) (Guillard et al., 2024), we identified that these patients more often reported somatosensory modulations of their tinnitus compared to control tinnitus patients without NITM. Likewise, patients with NITM more often reported increase of their tinnitus following physical activity (light exercise to intense workout). Additionally, patients displaying Tinnitus Increases After Naps (TIAN) more often reported modulations of their tinnitus by stress, sounds and sleep, suggesting a potential global hypersensitivity of the tinnitus percept of this subgroup to external stimuli. To the best of the authors' knowledge, no longitudinal polysomnographic exploration has been led to better understand these nap-induced tinnitus modulations (NITM).

There are several potential physiological events that can happen during naps that could explain NITM. A first possibility are bruxism events. It has been shown that tinnitus is significantly higher in the bruxism population (Camparis et al., 2005; Fernandes et al., 2014). One could hypothesize that bruxism events occurring during naps could constitute trigeminal nerve stimulations that could result in a somatosensory modulation of tinnitus upon awakening. Similarly, changes in cervical and masticatory muscle tonus, associated with long lasting inclined positions during sleep could be another potential somatosensory source of modulation explaining NITM. Last, one could argue that sleep apnea and snoring could be the source of NITM. Both of which have been shown to be related to tinnitus (Koo and Hwang, 2017; Lu et al., 2022) and both implicate the tensor veli palatini, another trigeminally innervated muscle.

In a previous study, we investigated a subgroup of TIAN patients : sleep-intermittent tinnitus (SIT) patients (i.e. patient who occasionally describe a total disappearance of the tinnitus after night sleep) (Guillard et al., 2023b). We compared, over a whole night of polysomnographic exploration, the sleep characteristics of a sample of this population to controls (tinnitus patients without any sleep-induced modulation). We did not identify any differences between these two populations with regard to sleep apnea, snoring, bruxism or sleep position during the night. We did not find any correlation between these variables and night sleep-induced tinnitus modulation either. Yet a strong limitation for the calculation of these correlations was the fact we only studied one-night of measurement per patient. A longitudinal set of measurements for each patient would appear necessary to be able to conclude upon what physiological mechanism correlates with sleep-induced tinnitus modulations. Indeed, in a seven-night longitudinal case-report of a SIT patient (supplementary material: Appendix), we identified that snoring weakly correlated with night-induced tinnitus modulations.

The present study was aimed at addressing the methodological limitations of our previous study by identifying the physiological correlates of NITM. In the present study, we primarily aimed at more specifically testing if NITM are linked to somatosensory modulations of tinnitus. In a subsequent study, we will analyze if NITM are related to central or autonomic changes. To achieve the goal of the present study, longitudinal assessments of tinnitus were performed through 6 naps per patient in a cohort of TIAN patients, along with a large set of audiologic, kinesiologic and polysomnographic explorations.

Such an experimental design has several theoretical methodological benefits.

First, it focuses on a homogeneous sub-phenotype of patients that share a least a similar pattern of tinnitus modulation. Indeed current guidelines in tinnitus research advocate for a comprehensive exploration of homogeneous subgroups of patients (Genitsaridi, 2021; Genitsaridi et al., 2020; Guillard et al., 2023a).

Second, the NITM in TIAN patients provides the unique advantage of inducing prolonged and stable modulations of tinnitus, facilitating extended post-nap assessments in the tinnitus-modulated state.

Lastly, the collection of longitudinal data from individual patients allows for the exploration of intra-individual correlations between tinnitus modulation and various physiological parameters, in a setup in which each patient is its own control.

## 2. Materials and Methods:

### 2.1. Participants :

One cohort of 37 TIAN patients (mean ± standard deviation age: 55.24 ± 10.70, 27 men, 10 women) were recruited to participate for two days of polysomnography examination at Hôtel-Dieu hospital sleep department (VIFASOM) in Paris, France. Most of these patients were recruited by ENT M.D. (A. L.), whereas some enrolled in the study spontaneously after a communication was made to social media platforms with the help of a French patient tinnitus association (France-Acouphènes) and the mutual help digital community Siopi (Guillard et al., 2023).

Inclusion criteria were: subjects aged between 18 and 75, reporting noticeable modulations of tinnitus following naps which can be masked by 85 dB broadband white noise, affiliated to the French social insurance and having signed informed consent. Non-inclusion criteria were: change in tinnitus treatment in the last 3 months, patients under guardianship or with important cognitive impairment or not able to do naps easily, or having been diagnosed with Ménière disease, acoustic schwannoma, severe hyperacusis or pulsatile tinnitus; as well as patients who reported epilepsy seizures, non-stabilized metabolic or cardiovascular conditions, or superficial scalp wounds or latex allergy and at last subjects pregnant or breastfeeding. Demographic and symptomatologic characteristics of the sample are presented in tables 1 and 2.

### 2.2. Ethics :

All patients gave their informed consent to participate to the present clinical trial, which received approval from the local ethical committee (CPP Idf3 2022-A00197-36). Before submission to the ethical committee, this protocol has also been validated by the patients' association board committee of France-Acouphènes. It was approved to take participants' best interests into consideration. Participants were compensated 100 euros for their participation and transportation fees were reimbursed up to 100 euros.

### 2.3. Clinical assessment :

2.3.1. Anamnesis case report form:

The participants completed a comprehensive Case Report Form (CRF). This CRF included a comprehensive description of tinnitus and its somatosensory modulations, history of comorbidities and associated pain symptoms (jaw, cervical area and headaches). CRF also included screening questions on sleep disorders and how tinnitus was modulated by night sleep and naps, and by sleep positions. Last, it included Visual Numerical Scales on tinnitus Loudness and Intrusiveness (VNS-L and VNS-I), following the methodological advice of the Comit'Id consensus (Hall et al., 2018). Finally, the Tinnitus Handicap Inventory (THI) was administered. The THI is a 25-item questionnaire yielding a score between 0 and 100 classically used in tinnitus clinical studies to assess the impact of tinnitus on quality of life (Newman et al., 1996). The main clinical characteristics of the sample are presented in Tables 1 and 2.

2.3.2. Protocol:

*Napping*

All participants were asked to perform six nap attempt cycles over two days in VIFASOM sleep lab. Each day had a similar planning. Minimal duration between the two days of participation at the hospital was of three days. Patients were advised not to drink alcohol nor coffee the days before and the days of the participation to the sleep lab.

Participants were first equipped with polysomnographic material (as described hereafter in section 2.2.3), then they went through a first nap attempt cycle in the morning starting 11am. Then came lunch, during which a selection of several menus with similar nutritional properties were proposed. The afternoon started by a second nap attempt cycle near 2pm, then a one-hour break, then a last nap attempt cycle near 4pm.

7 patients among the 37 recruited TIAN patients only participated to one day of clinical assessment, resulting in only 3 naps attempts performed for each of them.

Each nap attempt cycle was similar. A set of examinations were performed in the order hereafter before the nap, then in mirror order after the nap, as illustrated in Figure 1 (Pre-nap and Post-nap order of examinations reversed).

Napping occurred in an EEG quiet room. If the participant fell asleep, nap was interrupted after a maximum 15 min of N2 and/or N3 sleep to the best of the experimenter's ability to score sleep in real time. Participants were awoken by a gradual light bulb lighting over 1 min to minimize brutal awakening. The nap attempt stopped in all cases after 45 min, whether the participant fell asleep for less than 15 min or not. After the end of the nap attempt, participants performed the same set of examinations, in mirror order, so that the same time distance between examination and sleep was preserved before and after the nap (e.g. EEG resting state was the last examination before the nap and the first one upon awakening).

Polysomnographic recordings were performed using four Nox A1 devices (Resmed, San Diego, California, United-States). The sampling rates of the four devices were 250 Hz. EEG channels were positioned as usual at scalp locations F3, F4, C3, C4, O1, O2, and referenced to the opposite mastoid M1 and M2 according to the 10-20 system for electroencephalography (EEG) positioning (Jurcak et al., 2007). Two electrooculography (EOG) channels were positioned diagonally: one externally above one eye and the other externally under the other eye. as well as seven electromyography (EMG) sensors, two placed on the right and left suprahyoid muscles with one on the chin as reference, as well as two bipolar electrodes to measure left and right masseter activity, each referenced to the chin as well.

All EEG, EMG and EOG electrodes were gold cup electrodes filled with conductive paste (Weaver and Company - Ten20 Conductive Neurodiagnostic, Aurora, Colorado, United States), except for masseter bipolar electrodes which were consumable gel electrodes (Medtronics - Kendall - hydrogel ECG electrodes H92SG, Watford, United Kingdom). Epiderm was initially cleaned with alcoholic solution (Alcohol 70% vol.) and then dried before conductive paste application and electrode positioning. Electrodes were held in place using adhesive conductive cream (Natus Genuine Grass EC2 electrode cream, Middleton, Wisconsin, United States) on the top and adhesive band (Mölnlycke - Mefix - Self-adhesive fabric Gothenburg, Sweden).

Two electrocardiography (ECG) sensors were placed in left arm/right leg positions. Single-use humid electrodes were used as reference (on the forehead) and for ECG measurement (Medtronics - Kendall - hydrogel ECG electrodes H92SG, Watford, United Kingdom).

The Nox A1 units are equipped with additional sensors: accelerometers in the 3 dimensions, activity measurement, angular position of the device, dB audio measurement.

Sleep apnea was assessed by means of two plethysmography belts on the thorax and abdominal regions. Furthermore, oximetry and pulse were measured continuously by NONIN 3150 bluetooth oximeter (Nonin: Plymouth, Minnesota, United States) and nasal airflow was measured by a nasal cannula connected to the pressure sensor of the polysomnographic NOX A1 device. Naps were recorded with an infrared video camera.

Polysomnographic recordings were scored for sleep stages and sleep apnea following the ASSM recommendations by two experienced polysomnography technologists. For sleep scoring, a consensus scoring session was performed at the end of the 2 independent scorings to agree on epochs in which scorers had responded differently. For apnea scoring, the second scoring, which detailed the type of apnea events (hypopnea, central apnea, obstructive apnea), is presented in supplementary material (Supplementary Table 3) with its correlations with NITM. For bruxism scoring, automatic scoring was first performed with the Tinnitus-n-sleep toolbox (https://github.com/lkorczowski/Tinnitus-n-Sleep) and then reviewed manually. Bruxism scoring procedure for bursts and episodes for both automatic and manual scoring followed the scoring rules advised by (Lavigne et al., 1996). This toolbox enables to automatically label EMG jaw muscles signals for EMG bursts and scores bruxism events accordingly, combining successive burst into episodes following the recommendations given in (Carra et al., 2015; Lavigne et al., 1996). This task was performed using suprahyoid EMG channels while rejecting artifacts provoked by body movements (measured by the accelerometer embedded on the NOX A1), and faulty electrode contact with the skin (measured by electrode impedance). In addition to the analysis on the overall events, a more specific analysis was performed and is presented in the supplementary material.

Snoring events were manually scored from the decibel channel of the Nox devices. Cumulated duration and average snoring decibels were calculated from this scoring. Sleep apnea events were manually scored according to ASSM recommendations from air cannula airflow measurements, oxygen desaturation events (SPO2) and plethysmography measurements. SPO2 oxygen desaturation events and position assessment (supine, right, left, prone) were labeled automatically during the nap by the Noxturnal software (v6, Resmed, San Diego, California, United-States) respectively from the SPO2 signal and the accelerometers measurements, according to ASSM recommendations.

When a part of more than 60 seconds (2 epoch) of a nap had a missing signal, due to faulty or malfunctioning sensors, the feature it would enable to calculate was not evaluated for this nap. For example, if during a nap, the SPO2 sensor malfunctioned or fell from the finger for more than 60s, the number of desaturations was not assessed. This resulted in having a different number of data points for each test for each variable.

*Auditory assessments*

First a *tonal audiometry* at 250Hz, 500Hz, 750Hz, 1kHz, 1.5kHz, 2kHz, 3kHz, 4kHz, 6kHz and 8kHz auditory thresholds was collected for each ear with a descending procedure (Carhart and Jerger, 1959). Koalys online digital audiometer was used to acquire these audiograms, using Telephonics TDH-39 with noise-attenuating shells (Thai-Van et al., 2023).

*Multifrequency impedancemetry*, also called wideband tympanometry, was measured twice for each ear with a TITAN device (Interacoustics, Middelfart, Denmark) (Hein et al., 2017).

*Tubomanometry* (TMM) is an Eustachian tube (ET) function testing method. The principle of TMM is to deliver defined pressures of 30, 40 and 50 mbar to the epipharynx through a nasal applicator. Swallowing triggers the opening of the cartilaginous part of the ET and at the same time temporarily seals the epipharynx towards the pharynx. If the ET opens during swallowing, the defined pressure applied to the epipharynx is transmitted to the middle ear. Pressure curves of the epipharynx and the ear canal are displayed on the monitor of the TMM device, and various measuring values are calculated (Schröder et al., 2015). In the present study, TMM measurements were acquired three times before and after each nap with a TMM device from DTF Medical (Saint-Etienne, France).

*Physical examination*

*Neck mobility* was assessed through the Neckcare device (English et al., 2023), by recording range of motion of the neck in the three rotation directions, measured as angles in degrees. 3 measures were taken for each direction of rotation through an assisted and automatized procedure.

*Muscular trigger point tests* were then performed to assess masticatory and neck muscular sensitivity and its relation to tinnitus. These examinations were mainly realized by V.P. who had a physiotherapist background, who trained the other experimenters to perform the tests whenever he was not available. It consisted in the experimenter applying a defined and standard amount of pressure over the trigger points of a set of masticatory and neck muscles and asking the patient two questions. First question was to rate on a scale from 0 to 10 how much pain this pressure would induce, and then second to describe if it would elicit either an increase (encoded as +1), a decrease (encoded as -1) or no effect on his/her tinnitus loudness (encoded as 0) to test somatosensory modulation. Muscles were tested on the left and right side separately in the following order: trapezius muscles, sternocleidomastoid muscles, masseter muscles, temporalis muscles, splenius capitis muscles. All answers were collected in real time by experimenters on a dedicated Psychopy python interface.

An *adapted Spurling test* to assess pain elicited by compression of the cervical facet joints was applied on the left and right side (Michiels et al., 2015). Again, for each test, patients were asked if this maneuver would modify tinnitus loudness.

Next examinations were a set of *assessments of jaw mobility*. Range of motion of the jaw were measured in millimeters by a ruler for jaw maximum opening and laterotrusion to the right and left. Last, a jaw jerk test was performed (Bishop et al., 1984). Again, for each test, patients were asked if the performed movements would modify tinnitus perceived loudness.

### Tinnitus momentary assessment (TMA)

Then, except before the first nap attempt cycle, the patients were left in autonomy to perform TMA. TMA consisted first in reporting VNS-L and VNS-I. Then, a minimum masking level (MML) procedure was performed to estimate tinnitus loudness. MML was assessed with a MP4 device (Mobile Gear MP4 player, On-EARZ) and intra-auricular JBL Wired Headset (Harman International Industries). The masking stimulation was a broadband white noise (20 Hz to 10 kHz). MML procedure starting without any stimulation and then rising progressively the masking stimulation intensity up to the level tinnitus was masked. The exact instruction given to the participants was: "stop at the volume where you cannot discriminate your tinnitus perception from the sound stimulus, without actively focusing to do so". This procedure for MML assessment was performed and reported three consecutive times in order to ensure stability of the measurement. This set of procedures and how to fill their answers on the dedicated Psychopy python interface was taught to the patients just before the first nap attempt.

Last, patients went through an *EEG resting state measurement*. After a last verification of the quality of the signal, the advice given to the patient was to stare at a cross on the screen of a computer, staying still and without thinking about anything in particular for 90 seconds, and then when a bip sound would occur, to close their eyes and remain still without thinking about anything in particular until a second bip sound would end the EEG resting-state measurement.

During the period of time between the two days of participation, participants were asked to fill two short questionnaires one before going to bed and one shortly after awakening. These two questionnaires assessed tinnitus loudness and its associated intrusiveness using visual analog scales (respectively VNS-L and VNS-I), as well as Minimum Masking Levels (MML) of their tinnitus (with the method described in 2.3.4). These home TMA measurements were then included for the Z-scoring procedure described hereafter in section 2.5. This procedure was made to ensure we captured a better estimate of the variability of the distribution of tinnitus intensities experienced by the patients. Indeed, it could happen that the tinnitus intensity levels sampled during the two days at the hospital reflected only part of the variability of the real distribution. For example, it could happen if a patient arrived at the hospital in the morning with an almost maximal intensity on both days. In addition, the morning short questionnaire included a sleep agenda set of questions. Patients were asked to estimate the time at which they fell asleep and woke up, as well as to indicate whether there were any nocturnal awakenings or other notable events during the night.

A tinnitus matching procedure was performed to assess tinnitus pitch characteristics at the end of the first or the second day of participation depending on participant availability.

### 2.3.4. Experimenters blinding:

In order to limit bias, the experimenters were blinded to the variations of tinnitus perceived by patients over the course of the day. To ensure this, from the start of the day in hospital to the end of the last third of the day's nap cycle, participants reported their TMA autonomously on a custom-designed Psychopy interface (Peirce, 2007).

### 2.4. Preprocessing :

### 2.4.1. Assigning side variables according to tinnitus side :

Some of the variables used in this study are lateralized (for example left or right muscle, left or right hearing threshold) and so was tinnitus (left, right or bilateral). It appeared more valuable to measure the variables of interest relative to the tinnitus side. This is why we calculated metrics as "ipsilateral" or "contralateral" to the participant's tinnitus side. If a participant reported tinnitus "inside the head" or "in both sides", we used the average of left and right measurements.

### 2.4.2. Somatosensory assessment variables computation:

The answers collected during muscular trigger point tests were processed before statistical analyses. First, each right or left measurement was converted to a tinnitus ipsilateral or contralateral measurement according to the procedure described above. Then cumulated scores were calculated as the sum of the records for each muscle (trapezius muscle, sternocleidomastoid muscle, masseter muscle, temporalis muscle, splenius capitis muscle). These sums were calculated separately for ipsilateral and contralateral measurements for both sensibility scales and somatosensory modulation records. It resulted in four different cumulated scores.

### 2.5. Statistical analyses:

Data were centralized on a common anonymized CSV (Comma Separated Value) file and analyzed using Python, pandas library (McKinney, n.d.) and Scipy library (Virtanen et al., 2020).

Before any correlation tests, a z-scoring procedure was applied for all features. This procedure was as follows: centering and standardization was applied for all measurements of a feature on the set of naps from each patient separately. For example, for snoring duration, for each patient, Z-scoring was applied on the distribution of snoring durations of all his/her naps, independently from one patient to the other. TMA were performed between the two days of participation at the hospital: TMA included MML, VNS-L and VNS-I, which were assessed every morning and every evening for at least three days per patient. These additional measurements were included in the Z-scoring procedure to have better distribution granularity on these important TMA features.

A first set of tests were conducted to evaluate how naps and pauses between naps affected tinnitus intensity measured by MML. Wilcoxon paired-tests were used to evaluate the overall effect of naps on MML as well as between each distribution of MML values at each sequential time point of the day.

A second statistical analysis was led to evaluate the absolute amount of variation between each paired point for each time point of the day. Again, Wilcoxon paired-tests were used between each distribution of absolute difference of MML values between each sequential time point of the day.

Cohen-d effect size calculations were systematically performed for each Wilcoxon tests.

Features concerning the polysomnographic events measured during the nap (referred hereafter as intra-nap measurements) were correlated with TMA variations over the nap. More precisely, these TMA variations consisted in the Z-transformed MML, VNS-L and VNS-I scores after the naps minus the scores before the naps. As the dataset consisted of sets of naps of patients, Spearman correlation tests were performed between intra-nap measurements and TMA for each set of naps for all patients. Pearson correlation statistics were then extracted and aggregated in one list for all patients. Fisher Z-transform was then applied to this list of statistics. Then a Stouffer combination method was utilized on the given list of Z-transformed statistics to derive a single p-value for correlation group-wise. In cases where a Z-score value in the list exceeded 5 or fell below -5, it was assigned the value of 5 or -5, respectively.

Regarding the audiologic and kinesiologic explorations conducted before and after the nap (referred to as extra-nap measurements), their features were correlated with TMA scores before and after the nap. However, as these features' values were collected both before and after the nap, a Pearson 2D correlation test was employed. The resulting 2D individual correlation statistics were then compiled into a single list. The subsequent steps followed a similar procedure as described above: Fisher Z-transform was applied to this list of correlation statistics, followed by the application of the Stouffer combination method to obtain a final p-value for correlation group-wise. If any Z-score in the list exceeded 5 or fell below -5, it was respectively assigned the value of 5 or -5.

Holm-Bonferroni method (Abdi, 2010) was applied to control for multiple statistical testing on correlation tests. Such control was applied for each TMA feature set of tests separately, as TMA measurements were highly correlated one to the other (high correlation between MML, VNS-L and VNS-I measurements).

Last, the effect of naps on extra-nap measurements was assessed using a Fisher combination of p-value method over the list of one-sided paired Wilcoxon tests p-values for each participant scores before and after each nap.

## 3. Results :

### 3.1. Global effect of naps on tinnitus :

Figure 2 displays the representation of the Z-scored MML variation before and after a nap on a total of 193 naps (4 missing values for MML measurements). Naps that fall above or very close to the diagonal line indicate no variation in MML between before and after the nap. The majority of naps, as depicted in the top left corner, demonstrate an increase in MML after the nap, compared to the value before. This observation was further supported by a comprehensive test conducted on the 195 paired-values of MML, revealing a significant result ($p < 0.001$, Cohen-d: 0.55). However, it is important to note that not all naps resulted in an increase in MML. 23% of naps actually led to a reduction in MML, with certain participants reporting different outcomes across their 6 naps.

A supplementary analysis investing the potential relation between tinnitus severity and the amplitude of NITM was performed and is reported in the supplementary material, notably in Supplementary Figure 1.

*3.2. Effect of consecutive naps over the day on tinnitus :*

Figure 3 illustrates the changes in the MML measurements distributions throughout the day, encompassing all the days of participation from all participants. All three naps across the day elicited a significant increase of MML after the nap compared to the value before the nap (first nap : $p < 0.001$, Cohen-d : 0.59 , second nap : $p < 0.001$, Cohen-d : 0.59, third nap : $p < 0.001$, Cohen-d : 0.47). The first break of the day (2 hours during which patients had their lunch) elicited a significant reduction of tinnitus ($p < 0.001$, Cohen-d : 0.7), as well as the second break (1 hour, no systematic food intake, $p < 0.05$, Cohen-d : 0.34).

Figure 4 illustrates the absolute magnitude of MML variation between paired MML values for each time interval of the day. As the day progressed, there was a gradual decrease in the impact of naps. We observed a notable decrease in the absolute paired-difference of MML between before and after the first nap and before and after the second nap ($p < 0.05$, Cohen-d: 0.41). Similarly, we also observed a significant decrease in magnitude between the first and the third nap ($p < 0.01$, Cohen-d: 0.49). However, there was no significant reduction observed between the second and third nap of the day.

*3.3. Effect of naps on extra-nap measurements :*

Effect of naps was tested on extra-nap measurements. Results are reported in supplementary material, in Supplementary Table 4.

*3.3. Correlations of TMA metrics with intra-nap measurements :*

Table 3 displays the outcome of the correlation analysis conducted on the intra-nap measurements and TMA measurements. We provide further information regarding each measurement subsequently.

3.3.1 Correlations with nap sleep duration:

Light sleep (N2 stage) was reached for at least 1 epoch (30s) in 169 out of the 197 nap attempts performed (85, 7 %). Deep sleep (N3) was present in 21 out of the 195 naps (10 , 7%) performed. No participant reached REM sleep stage in any epoch during any of the nap attempts. After Holm-Bonferroni statistical correction, significant positive correlations were found between N2 duration during the nap and VNS-L variation ($p < 0.05$) and N2 and N3 durations during the nap and VNS-L variation ($p < 0.05$). We also performed a supplementary correlation test between the nap attempt total duration (independently of sleep duration) and VNS-L variation and we did not find significant, even without any correction for statistical testing.

3.3.2 Correlations with sleep breathing disorder metrics:

Different sleep breathing disorder metrics were examined in this study. Firstly, we studied the duration of snoring and the average sound intensity of snoring (in dB). The results revealed significant positive correlations between the duration of snoring and the variation in MML ($p < 0.001$), as well as between the average sound intensity of snoring and the variation in VNS-I ($p < 0.05$). To further validate the findings, a supplementary control test was conducted to assess the correlation between the general average sound in the room during the nap (in dB) and each TMA metric. However, even without applying any statistical correction, no significant correlation was observed.

Interestingly, in the initial questionnaire, 19 participants (51.35%) reported snoring, but during the naps recording, 33 participants (89.19%) exhibited more than 30 seconds of cumulative snoring during at least one of their naps. Likewise, in the initial questionnaire, 8 participants (24.32%) reported being diagnosed with sleep apnea. However, we did not include full night polysomnography or polygraphy in this protocol to confirm objectively these diagnoses. Besides, during the nap recordings, 34 participants (91.89%) exhibited 15 or more sleep apnea or hypopnea events per hour of sleep during at least one of their naps.

Another important objective of this experiment was to determine the correlation between the variation in MML and the count of apnea and hypopnea events throughout the nap. While a significant positive correlation was observed between the variation in VNS-I and the count of apnea and hypopnea events ($p < 0.001$), the absence of its correlation with MML and VNS-L measurements, even without statistical correction, questions the direct implication of sleep apnea in NITM. Moreover, no significative correlations were observed between TMA and SPO2 desaturation count.

More detailed analyses on the implication of sleep breathing disorder in NITM are presented in the Supplementary Table 1. In this table, it is worth noting that when cumulating the durations of snoring events and sleep apnea events during the nap, reflecting the thus the total number of sleep breathing disorder events, in the nap, we observe, without statistical corrections, a significative correlation between this metric and all three TMA metrics (MML, VNS-L and VNS-I, $p < 0.01$).

### 3.3.3 Effect of Sleep position on tinnitus modulation :

Out of the 37 participants who took part in this study, a mere 2 participants (5.41%) stated that taking a nap while seated tended to exacerbate their tinnitus. On the other hand, a majority of 31 participants (83.78%) reported that their tinnitus remained unaffected by their sleeping position. This initial finding was further supported by the absence of any significant correlation between the sleep position during the nap and any of the TMA metrics. Detailed analysis of the head position during the nap, as determined through manual scoring of video camera footage, can be found in the supplementary material 2.

### 3.3.4 Effect of bruxism on tinnitus modulation :

Out of the 37 individuals who took part in this study, merely 7 participants (18.92%) disclosed that they had been diagnosed with sleep bruxism. Throughout the nap recordings, 6 participants experienced 5 or more episodes of sleep bruxism in at least one of their naps. There was no observed correlation between the number of sleep bruxism episodes and any of the TMA metrics.

### 3.4. Correlations of TMA metrics with extra-nap measurements :

No statistically significant correlation was found between any of the extra-naps measurements (trigger point tests measurements, neck and jaw mobility, auditory thresholds nearest to tinnitus frequency, average tone loss, multifrequency impedancemetry measurements) and any of the TMA metrics after statistical correction. Tubomanometry measurements appeared unexploitable for analysis due to poor signal quality and test reliability.

Correlations of TMA and extra-nap measurements are presented in greater detail with all individual items instead of aggregated scores in supplementary material table 2 to this study. Such a supplementary analysis was purely exploratory and thus no statistical corrections were applied.

## 4. Discussion

This study is the first to demonstrate and quantify in a controlled clinical environment that naps can increase tinnitus. It gives objective ground to the experiences frequently reported by tinnitus patients seeking medical advice.

This study is indeed the first to investigate correlates of Naps-Induced Tinnitus Modulations (NITM) in the Tinnitus Increases After Naps (TIAN) population. It has successfully identified potential factors that correlate with NITM, such as snoring and sleep breathing disorder. Our study involved a longitudinal analysis of a sample of patients, who underwent six nap attempts over two days in a sleep laboratory. The results revealed that NITM is not a consistent phenomenon, as we observed different variations of tinnitus in the same individuals during the six nap attempts. However, we observed at a group level that each nap, regardless of the time of day, led to a significant increase in mean tinnitus loudness. Interestingly, these increases were temporary, as we also remarked significant decreases in tinnitus loudness between naps.

We believe that the introduction of longitudinal measurements of NITM in the TIAN population offers several methodological advantages. Firstly, it allows for a comparison of participants with themselves, this allows for more statistically robust pairwise comparisons by eliminating many confounding factors, since the patient is his or her own control. It is important to emphasize that the majority of previous protocols have focused on comparing a tinnitus sample to matched controls in order to identify physiological correlates of tinnitus. (Kara et al., 2020; Lanting et al., 2009; Schaette and McAlpine, 2011; Weisz et al., 2005). But such comparison between different groups of participants brings unpaired measurements that can hardly be perfectly matched, a key example being different patterns of hearing loss, or unknown amounts of auditory nerve deafferentation (Adjamian et al., 2014, 2009). Indeed, a variety of studies have already tried to compare tinnitus patients to themselves in a tinnitus modulated state, often through sound stimulation. Examples are residual inhibition (Gentil, 2016), Zwicker tone (Mohan et al., 2020), mechanical stimulation of the middle ear (Job et al., 2016). Such modulation techniques have at least two limitations: first, the modulation is very short, not leaving enough time to perform time-consuming audiologic tests. Second, as the technique used to induce the modulation is sound based, any test aiming at probing the auditory system suffers the bias of the sound stimulation. NITM solves these limitations as patients of the TIAN population often report that NITM are long and stable modulations of tinnitus.

Additionally, the transition from a normal tinnitus state to a modulated state during naps is relatively short (15 minutes of sleep). Lastly, as demonstrated in this study, naps have minimal physiological effects apart from NITM. As described in the Supplementary material, although non-significative, a tendency of modification of the resonance frequency of multifrequency impedancemetry was observed, accompanying an associated tendency of modification of the audiogram. It was shown that multifrequency impedancemetry did not correlate with NITM with our 2D correlation test. Such a reduction in resonance frequency of multifrequency impedancemetry following naps could be explained by a residual effect of middle-ear pressure elevation during sleep (Thom et al., 2015; Tideholm et al., 1999). Interestingly, we conducted a supplementary test between the difference of resonance frequency between before and after the nap and the variation of the TMA metrics over the nap and we found a significative negative correlation with all three TMA metrics (p < 0.01, without correction). Overall, NITM provides a valuable framework for studying and identifying the physiological factors associated with tinnitus. This study contributes to the understanding of NITM and its potential links to somatosensory triggers, furthering the knowledge in this field.

*NITM and the somatosensory hypothesis :*

The motivation behind this study was to identify the physiological correlates of NITM, and more specifically to test if NITM were a hidden form of somatosensory modulation (Levine, 2013; Levine and Oron, 2015; Ralli et al., 2016). We systematically tested a variety of physiological correlates that could be involved in generating somatosensory modulations of tinnitus. We performed some of these assessments during the naps: sleep apnea, snoring, bruxism and sleep position. Other assessments were tested before and after the nap along with TMA: neck and jaw mobility, neck and jaw muscle sensitivity and susceptibility to elicit somatosensory modulations and middle-ear multifrequency impedancemetry.

Somatosensory modulations of tinnitus are believed to be triggered by somatosensory information transmitted by the muscles and joints innervated by the mandibular branch of the trigeminal nerve to the cochlear nucleus via the caudal spinal part of the trigeminal complex (Sp5) and by dorsal root ganglions (Shore et al., 2007, 2016). If NITM were indeed a hidden form of somatosensory modulation, it would be expected that naps would alter the somatosensory signal transmitted by at least one muscle or joint structure innervated by the trigeminal nerve. This change in the afferent somatosensory signal would consequently lead to a corresponding change in tinnitus loudness.

The present results did not enable to identify any connections between the overall cumulated somatosensory information transmitted by the jaw and neck tested muscles and the modulation of tinnitus by naps. Indeed, no significant correlations were observed between TMA and the cumulative scores of trigger point sensitivity or somatosensory modulations of tinnitus. Further analysis, as described in the supplementary material, showed that even when examining each muscle individually, only the sensitivity changes in the contralateral trapezoid muscle and contralateral sternocleidomastoid muscle exhibited a significant correlation (without statistical corrections) with TMA.

Moreover, we did not observe a link between temporomandibular joint and masticatory muscles and NITM. Indeed, neither jaw mobility nor bruxism episode counts significantly correlated with TMA.

However, the mandibular branch of the trigeminal nerve does not solely innervate the masticatory muscles. It also innervates the tensor veli palatini (TVP) and tensor tympani (TT). Previous studies have shown that they could play a role in tinnitus generation and/or modulation, notably in the case of the tensor tympani syndrome or in case of acoustic shock (Londero et al., 2017; Westcott et al., 2013). A review conducted by Norena et al. proposed that the excessive use of the TT muscle can give rise to a chronic neurogenic inflammation. This inflammation would lead to the continuous transmission of somatosensory information to the trigeminal complex (acoustic shock model) (Noreña et al., 2018).

*Tensor veli palatini overuse hypothesis :*

In the TIAN sample recruited for the present study, we observed an abnormally high prevalence (90%) of snoring and of at least mild sleep apnea (apnea-hypopnea index of 15 or more) in at least one the naps of the participants. Additionally, we found significant correlations between snoring duration and MML variation, between average snoring sound intensity and VNS-I and between apnea count and VNS-I. In a previous 7-night polysomnographic case report (see supplementary material of (Guillard et al., 2023b)), we also had observed a positive correlation between overnight tinnitus intensity variation and percentage of snoring duration over the night sleep duration. These findings provide support for the notion that a similar model to the acoustic shock model could be applied to the TVP instead of the TT for the TIAN population. Sleep apnea and snoring are both very often related to TVP tonus dysregulations during sleep (Patel et al., 2018; Zhu et al., 2022). In fact, some studies suggest that people with snoring and/or obstructive sleep apnea syndrome (OSA) have by nature a narrower pharyngeal airway section (Haponik et al., 1983). Consequently, this would result in a higher tonic tonus of the TVP in the awake state to maintain airway patency. A study conducted by Mezzanotte et al. (Mezzanotte et al., 1996) also suggested this possibility, showing that apneic patients display a more significant decrease in electromyographic tonus upon sleep onset compared to healthy controls. Consequently, this would lead to an increase in somatosensory information transmitted by the muscle to the trigeminal complex, thereby altering its structure. Evidence of this phenomenon has been demonstrated in rodent models of sleep apnea, where an increase in noradrenergic terminals in the Sp5 region of the trigeminal complex was observed (Mody et al., 2011). In this model, sleep occurrences, such as naps, would constitute triggers for changes in TVP tonus. This would result in changes of somatosensory information afferent signal and as a result of tinnitus perception through somatosensory modulation of the auditory signal in the central auditory pathway.

Additionally, in alignment with the acoustic shock model, chronic apnea and/or chronic snoring could trigger a chronic neurogenic inflammation, reactivated night after night. This possibility is strongly supported by several reviews and a meta-analysis which concluded that obstructive sleep apnea was associated with systemic inflammation (Bergeron et al., 2005; Gnoni et al., 2022; Nadeem et al., 2013; Unnikrishnan et al., 2015). Inflammation is as well present locally as a study reported proliferation of receptors for nociceptive agent substance P and calcitonin gene-related peptide (CGRP), in the palatal mucosa of apneic patients and chronic snorers (Friberg et al., 1997).

Furthermore, snoring activity during sleep could constitute abnormal afferent proprioceptive signals that could contribute to somatosensory modulation of tinnitus. Indeed, one definition of simple snoring consists of a fluttering sound created by the vibrations of pharyngeal tissues (Deary et al., 2014). Such vibrations would generate a proprioceptive excitation of the TVP and the TT, then transmitted to the trigeminal complex. This proprioceptive disturbance could in itself be a source of modulation of tinnitus as a past study has hypothesized that proprioceptive signal disruption within the middle ear by low frequency vibrations in the range of snoring-induced vibrations could trigger tinnitus (Job et al., 2016).

*Alternative possibilities :*

Another alternative explanation for the observed association between sleep breathing disorder and NITM could be the potential impact of both hypoxia or hypercapnia induced by these events on NITM. Recent research has indicated a link between sleep apnea and hearing loss (Ekin et al., 2016). Similarly, sleep apnea appears to be highly co-prevalent with tinnitus (Koo and Hwang, 2017). It would be plausible to consider that following a nap with sleep apnea episodes, patients may experience a temporary deterioration in their hearing, leading to an increase in tinnitus intensity. However, this particular scenario is unlikely based on the findings of the present study. We did not find any correlations between the number of SPO2 desaturation events and TMA. Furthermore, naps did not appear to have any impact on hearing thresholds.

Then, another explanation for the observed relationship between the duration of snoring and MML, as well as the average sound intensity of snoring and VNS-I, could be attributed to the potential increase in tinnitus intensity caused by exposure to sound during the nap. We there hypothesized that patients may unknowingly subject themselves to some form of acoustic trauma/shock, resulting in an elevation of their tinnitus loudness upon waking up. However, it is important to note that this possibility is unlikely, as the control tests examining the correlation between the average sound intensity in the room during the nap and TMA did not yield statistically significant results, even without applying any corrections. Yet, the fact that snoring sounds do not generally induce the awakening of the snorer may reflect they are specifically gated during sleep. The network involved in gating sounds during sleep probably overlaps with the noise cancelling network described by Rauschecker and colleagues (Rauschecker et al., 2010). It may notably involve the pregenual and dorsal anterior cingulate cortex regions and the auditory cortex that were identified by Vanneste and colleagues to play the role of a tinnitus on-off switch (Vanneste et al., 2024).

*Cervical and body position implication in NITM:*

Finally, we examined the hypothesis of the potential involvement of the cervical area in NITM. Previous research has demonstrated that the second cervical dorsal root ganglion connects to the cochlear nucleus in animal models (Zhan et al., 2006) suggesting that it could play a role in somatosensory modulation of tinnitus (Wu et al., 2015). Some studies have proposed that certain forms of tinnitus may be closely related to cervical spine dysfunctions in the upper cervical region, such forms of tinnitus often referred to as cervicogenic somatosensory tinnitus (Koning, 2020a, 2020b; Michiels et al., 2015; S. Michiels et al., 2016; Oostendorp et al., 2016). Several therapeutic interventions on the upper cervical area have been proposed to alleviate tinnitus (Sarah Michiels et al., 2016; S. Michiels et al., 2016; Vanneste et al., 2010).

Previous studies have suggested that NITM may occur more frequently when individuals sleep in a seated position (Levine, 2013; Levine and Oron, 2015; Ralli et al., 2016). Sleeping in the seated position may induce cervical instability and prolonged periods of bad neck posture.

However, the findings of the present study indicate that the involvement of the upper cervical spine or body position in NITM is unlikely based on the observed results. There were no significant correlations found between neck mobility, body position, and TMA. Additionally, specific trigger point tests conducted on the splenius capitis muscles did not show any correlation with TMA. When examining the correlation between head position during sleep and TMA, only a significant correlation without statistical corrections was found between head tilting towards the same side and opposite side of tinnitus and VNS-L (cf supplementary material). Lastly, only 2 patients (5.41%) reported an increase in tinnitus intensity when sleeping in a seated position.

*Sleep occurrence and NITM:*

An essential question that arises concerning NITM is whether sleep is a prerequisite for its manifestation. The current study provides limited evidence suggesting that sleep is necessary for NITM to take place. To test this question directly, we could have made a comparison between nap attempts with sleep and those without sleep by examining TMA. However, out of the 197 recorded naps, only 10 nap attempts did not involve any sleep, not even the N1 light sleep stage limiting the possibility of comparison. When comparing these 10 naps to the others that included at least one epoch of sleep, we found no significant differences in terms of TMA amplitude.

Nevertheless, indirect tests bring more insights on this question. Significant correlations could be drawn between N2 sleep duration and VNS-L and cumulated duration of N2 and N3 sleep stages and VNS-L. To control if these correlations were not due to the duration spent in a dark and silent room and not because of sleep, we tried to correlate nap attempt duration length and TMA metrics. No significant correlations were found. This suggests a specific dose effect of sleep or an effect of sleep depth.

Such a finding opens the door for a potential alternative scenario: NITM may be induced by central mechanisms that are directly linked to sleep-induced changes in brain functioning. Further explorations on electroencephalographic data acquired during the nap and during the resting states measurements before and after each nap are required to give more ground to this possibility.

It should be noted that in our previous study comparing the polysomnographic characteristics of SIT patients to matched controls and in a subsequent longitudinal case study of 7 nights of an SIT patient (Guillard et al., 2023b), we did observe a difference of sleep depth between SIT patients and matched tinnitus controls (more N2 sleep and less N3 and REM sleep in SIT patients). Moreover, we observed a negative correlation between REM sleep duration and overnight tinnitus intensity variation in the 7-night case report. However, we did not observe differences or correlations with night sleep duration neither in the comparative study nor in the case report.

*Audiological changes and NITM:*

Tinnitus is commonly associated with some form of hearing impairment, even though traditional audiograms may sometimes appear normal (Jafari et al., 2022). Recently significant loss of auditory nerve fibers were linked to presence of tinnitus in subjects presenting normal hearing audiograms (Vasilkov et al., 2023). One could hence hypothesize that NITM may be due to a transient hearing loss induced by naps.

Our observations tend to reject such a hypothesis as no significant correlation was observed between average tone loss and TMA. Detailed correlation tests between the audiogram features and TMA are reported in Supplementary Table 2. In these analyses, only a significant positive correlation (without statistical corrections) was observed between ipsilateral 4 kHz auditory thresholds and TMA metrics VNS-L and VNS-I. However, it is difficult to draw a direct connection between this correlation and the frequency characteristics of tinnitus in the sample, as the average tinnitus frequency was 7155 Hz, as shown in table 1. Additionally, it is worth mentioning that the audiogram characteristics remained unaffected by naps during the study. This finding suggests that either naps solely induce changes in the perception of tinnitus without affecting hearing abilities, or that potential changes in hearing ability are too subtle to be detected through modifications in the audiogram.

Besides, we conducted another supplementary analysis to assess if generally MML levels correlated with the degree of hearing loss. As displayed in supplementary figures 2 and 3 of the supplementary material, we observed significative correlations between average MML scores over all the naps (in dB) and both the average tone loss (Spearman correlation, N=37, $p < 0.05$) and the auditory thresholds of the audiogram closest to the tinnitus frequency and ipsilateral to the tinnitus Spearman correlation, N=37, $p < 0.01$).

*4.3. Limitations :*
.

One of the primary constraints of the current study was the presence of missing or unusable data, which hindered the analysis process. It is important to highlight that out of the 37 participants, 7 individuals were involved in the hospital study for just one day. Since three nap attempts were deemed insufficient to conduct individual correlation tests, this led to a smaller sample size for correlation analysis.

A second limitation of this study was associated with its exploratory nature. Being the first study of its kind to attempt to identify physiological correlates of NITM, numerous hypotheses were tested. However, the moderate sample size that could be recruited within the designated time frame (originally aiming at 40 participants) required statistical corrections for multiple testing, which led to the suppression of many seemingly significant results.

Another limitation was that the study did not assess whether NITM is influenced by circadian rhythms or homeostatic sleep pressure. The experimental design used in this study was unable to separate the effects of repeated naps throughout the day (and thus the progressive dissipation of homeostatic sleep pressure) from the circadian effect on NITM. There are multiple ways circadian rhythm could influence NITM. NITM could increase or could decrease in magnitude as the day progresses due to a circadian effect.

As shown in Figure 2, it was observed that the magnitude of NITM decreases over time. This observation suggests at least that it is unlikely that a circadian effect causes NITM to increase as the day progresses. On the other hand, it is plausible that the repetition of naps across the day progressively dissipated sleep pressure. Previous studies, notably in older subjects as in our sample, converge to show that doing a nap tend to increase sleep latency and occurrences of nocturnal awakenings the following night, which suggests an induced decrease in sleep pressure (Faraut et al., 2017; Saletin et al., 2017; Yoon et al., 2003). Moreover, a theoretical review recently proposed that sleep pressure could interfere with tinnitus perception (Milinski et al., 2022). In the present study, such a scenario is credible as we overall observed a progression from 8 failed nap attempts (no successful sleep onset) for the first attempt, 9 for the second and 17 for the third nap attempt.

However, within the experimental design of the present study, circadian and sleep pressure effects cannot be totally disentangled. As a consequence, it cannot be concluded whether the pattern observed is due to a circadian amplification of NITM in the morning, a decrease in NITM amplitude due to repeated naps (and thus due to sleep pressure modification), or a combination of both.

Furthermore, the results of this study revealed a minor flaw in the experimental design. The experimenters believed that NITM remained stable over time after a nap based on patients' reports. This led them to confidently perform audiological and kinesiological measurements post-nap, expecting the tinnitus modulated state to remain stable during these assessments. However, the present results showed a significant decrease in tinnitus loudness during the breaks between naps. Since the post-nap measurements of MML, VNS-L, and VNS-I were conducted shortly after participants woke up, it is possible that tinnitus decreased back to its baseline level before the end of the measurements. To address this issue and ensure the stability of NITM, a simple solution would have been to measure MML, VNS-L, and VNS-I both at the beginning and at the end of pre-nap and post-nap measurements.

Then another limitation of the present study is the lack of control procedures on the correlational tests conducted which hinders the ability to ensure independence among variables. For instance, the duration of nap sleep was found to be correlated with VNS-L, and before statistical corrections, such correlation was also present with VNS-I. However, it is important to note that apnea events count and durations of snoring are both variables that are correlated with durations of nap sleep. Therefore, when measuring either of these two variables, a portion of the observed correlation may actually be attributed to the relationship between nap sleep duration and TMA. To address this issue, more rigorous statistical procedures could be employed to eliminate the influence of independent variables on dependent ones. Yet, such additional procedures will require significantly more instances of nap attempts per patient.

Additionally, potentially useful information was not collected in the CRF initial questionnaire: body mass index (height and weight) that could have been important as a sleep apnea prevalence confounding factor, mental health questionnaires to assess anxiety, depression or mood.

### 4.4. Suggestions for future research:

As stated just above, future protocols trying to identify NITM physiological correlates should make TMA measurements at the beginning and the end of the pre-nap and post-nap measurements.

Another suggestion would be to increase the number of naps per patient, as six naps were maybe not enough to test whether or not individual correlations were significant. As the number of days one participant will be motivated to attend to the hospital for a non-interventional study is limited, a good suggestion would be to try to make it possible for participants to perform their nap attempts at home. It would enable performing these naps on a longer time scale, possibly under remote supervision of the experimenters. Such sleep remote monitoring that can be performed in autonomy by participants could involve the use of novel smart medical devices such as smart rings as the Oura ring (Oura Health Ltd., Oulu, Finland), the Circular Ring (Circular SAS, Paris, France) or EEG headsets such as the Dreem headset (Beacon Biosignals, Boston, USA).

Another suggestion would be to combine nap and night measurements in the same cohort of patients so as to test if naps and night induced tinnitus modulations are caused by the same physiological mechanisms.

Last, a subsequent study focusing on analyzing the EEG activity of participants collected within the present study (before, during and after the naps) is planned to address the alternative hypothesis of whether or not NITM are maybe not induced by somatosensory modulations of tinnitus but by central changes of brain activity induced by naps.

## 5. Conclusions

The current investigation is the first polysomnographic exploration of the TIAN population, aiming to elucidate the underlying mechanisms through which naps may influence tinnitus. We particularly emphasized the methodological design of this study to minimize potential sources of bias when evaluating the associations between NITM and potential physiological factors. Substantial increases in tinnitus loudness, as measured by MML, were observed following each nap session (one in the morning and two in the afternoon). The present study aimed at investigating whether or not NITM were a hidden form of somatosensory modulation. Present results tend to show it is unlikely that NITM could be a tinnitus somatosensory modulation triggered by the jaw or the neck. Instead, results suggest a potential correlation between TMA and nap sleep duration, sleep breathing disorders and especially snoring. Such observations suggest a potential involvement of the TVP muscle in NITM. A new hypothesis on how chronic snoring and sleep apnea could constitute a covert form of tinnitus somatosensory modulation is discussed. Further clinical investigations employing the same methodological framework are mandatory to validate these preliminary exploratory outcomes.

**Supplementary Materials:** The supplementary material "Exploratory correlational analyses" is available hereafter

**Author Contributions:** Conceptualization, R.G., D.L., B. F., S.M. and A.L.; methodology, R.G., D.L., S.M., B.F. and A.L.; software, R.G. and A.H.; validation, R.G., D.L. and A.L.; formal analysis, R.G. V.P. and A.H.; investigation, R.G., V.P, A.H., B.F., D.L and A.L.; data curation, R.G., V.P., A.H., M.P.; writing—original draft preparation, R.G.; writing—review and editing, R.G., M.C, D.L., B.F., S.M. and A.L.; supervision, R.G., M.C., D.L. and A.L.; project administration, R.G., A.L. and D.L.; funding acquisition, R.G. All authors have read and agreed to the published version of the manuscript.

**Funding:** This research was funded by Felicia and Jean-Jacques Lopez-Loreta Fondation grant to R.G..

**Institutional Review Board Statement:** The study was conducted according to the guidelines of the Declaration of Helsinki, and approved by the Institutional Review Board (or Ethics Committee) of CPP Iff3 under RIPH2G reference 22.00921.000121

**Informed Consent Statement:** Informed consent was obtained from all subjects involved in the study.

**Acknowledgements:** The authors would like to thank Mr Louis Korczowski for his contribution in the realization of the Tinnitus-n-Sleep github repository that was used for labeling bruxism events and for his role as founder of Siopi. Similarly, the authors would like to warmly thank France-Acouphènes and the Siopi mobile app team for their help to recruit subjects for the study. The authors would like to thank Mrs Charlotte Glabasnia Linck for her contribution of independently scoring a second time the sleep apnea events of the nap recordings, as presented in the supplementary table 3.

**Financial disclosure:** Damien Léger declares that in the past 5 years he has been employed as an investigator or a consultant by Actellion-Idorsia, the Agence Spatiale Européenne, Bioprojet,

iSommeil, ESAI, Jansenn, Jazz, Vanda, Merck, Philips, Rythm, Sanofi, Vitalaire, and Resmed. Robin Guillard declares that he is shareholder and president in Siopi SAS, and has a professional activity as independent as Robin Guillard EIRL.

The other authors did not declare financial interests.

**Non-financial disclosure:** None.

**Data availability statement:** The data underlying this article will be shared on reasonable request to the corresponding author.

**Use of generative AI and AI-assisted technologies in scientific writing**: None of the content of this manuscript was generated through a generative AI tool. The authors wrote the entire manuscript and then used an AI-assisted English stylistic improvement tool to improve the quality of English only in some sections of the manuscript ( https://ahrefs.com/writing-tools/sentence-rewriter ).

## References

References must be numbered in order of appearance in the text (including citations in tables and legends) and listed individually at the end of the manuscript. We recommend preparing the references with a bibliography software package, such as EndNote, ReferenceManager or Zotero to avoid typing mistakes and duplicated references. Include the digital object identifier (DOI) for all references where available.

Citations and references in the Supplementary Materials are permitted provided that they also appear in the reference list here.

Table 1, Sample characteristics for quantitative features (N =37)

| | Mean | Std | Min | Max |
|---|---|---|---|---|
| Age (in years) | 55.24 | 10.70 | 32 | 75 |
| Tinnitus Handicap Inventory (THI) | 48.76 | 20.43 | 14 | 98 |
| VNS scale Tinnitus loudness (0 to 10) | 6.24 | 1.83 | 3 | 10 |
| VNS scale Tinnitus intrusiveness (0 to 10) | 5.89 | 2.34 | 2 | 10 |
| Tinnitus main frequency | 7155.1 | 3312 | 275 | 12000 |
| *Abbreviations : VNS : Visual Numeric Scale* | | | | |

*Table 1 – Sample characteristics for quantitative features, N=37, Abbreviations : VNS : Visual Analog Scale*

## Table 2, Sample symptomatologic characterisation (N = 37)

| General tinnitus characteristics | Percentage |
|---|---|
| **Assumed cause of tinnitus*** | |
| Exposure to loud sounds | 21.62 % |
| Changes in hearing | 8.11 % |
| Exposure to a change in ambient pressure | 0.0 % |
| Flu, cold, or other infection | 2.7 % |
| Feeling of plugged ears | 18.92 % |
| Episode of stress | 54.05 % |
| Head trauma | 5.41 % |
| Trauma to the neck (e.g., whiplash) | 5.41 % |
| I don't know | 10.81 % |
| Other | 24.32 % |
| **Description of tinnitus sound*** | |
| I perceive only one sound | 24.32 % |
| I perceive several different sounds | 29.73 % |
| The overall tone is high | 67.57 % |
| The overall tone is medium | 5.41 % |
| The overall tone is low | 5.41 % |
| One sound I hear resembles a pure tone | 54.05 % |
| One sound I hear resembles a whistling | 5.41 % |
| One sound I hear resembles a buzzing | 21.62 % |
| One sound I hear seems melodious | 2.7 % |
| One sound I hear resembles crickets | 8.11 % |
| One sound I hear resembles a background noise | 18.92 % |
| One sound I hear resembles a hissing | 29.73 % |
| Other | 13.51 % |
| **Tinnitus localisation (most of the time)** | |
| On both sides? | 24.32 % |
| On both sides but more to the left? | 24.32 % |
| On both sides but more to the right? | 16.22 % |
| Exclusively to the right? | 13.51 % |
| Exclusively to the left? | 10.81 % |
| In your head? | 8.11 % |
| Other | 2.70 % |
| **Tinnitus onset** | |
| Sudden | 59.46 % |
| Progressive | 40.54 % |

| Tinnitus commorbidities | Percentage |
|---|---|
| **Tinnitus Duration** | |
| Between 6 months and 1 year | 10.81 % |
| Between 1 and 2 years | 2.7 % |
| Between 2 and 5 years | 48.65 % |
| Over 5 years | 37.84 % |
| **Self-presumed hearing loss** | |
| No | 29.73 % |
| Yes, a slight hearing loss | 35.14 % |
| Yes, a moderate hearing loss | 13.51 % |
| Yes, a significant hearing loss | 10.81 % |
| Don't know | 10.81 % |
| **Many everyday noises painful or bothersome** | |
| No | 54.05 % |
| Yes | 45.95 % |
| **Cervical History*** | |
| Cervical osteoarthritis | 27.03 % |
| History of whiplash or cervical trauma | 18.92 % |
| Herniated disc | 18.92 % |
| Cervical discopathy | 10.81 % |
| Scoliosis | 10.81 % |
| Kyphosis / kyphotic posture | 16.22 % |
| Surgical intervention at the cervical level | 0.0 % |
| No history to declare | 54.05 % |
| Other | 5.41 % |
| Don't know | 37.84 % |
| **Migraines or headaches** | |
| Almost never | 45.95 % |
| Rarely | 32.43 % |
| Sometimes | 16.22 % |
| Often | 13.51 % |
| Very often | 0.00 % |
| **Feeling of ear fullness or plugged ears** | |
| Almost never | 45.95 % |
| Rarely | 18.93 % |
| Sometimes | 18.93 % |
| Often | 8.11 % |
| Very often | 18.93 % |

| Tinnitus variation | Percentage |
|---|---|
| **Variation of Tinnitus Over Time*** | |
| From second to second without any reason | 10.81 % |
| Different intensities over different days | 72.97 % |
| Frequently louder upon waking | 51.35 % |
| Frequently louder in the evening | 21.62 % |
| No tinnitus variation or negligible | 0.0 % |
| Other | 5.41 % |
| **Variation of Tinnitus with Movements*** | |
| When you clench your teeth | 35.14 % |
| Mouth opening with resistance | 37.84 % |
| Jaw protrusion with resistance | 45.95 % |
| Jaw abduction with resistance (left) | 32.43 % |
| Jaw abduction with resistance (right) | 32.43 % |
| Head rotation (left or right) | 27.03 % |
| Head rotation with resistance (left or right) | 32.43 % |
| Lowering the head | 13.51 % |
| Lower the head with resistance | 35.14 % |
| Tilting the head backward | 10.81 % |
| Tilting the head backward with resistance | 24.32 % |
| None of the above | 40.54 % |
| **Loud sounds increase tinnitus** | |
| No | 48.65 % |
| Yes | 27.03 % |
| Don't know | 24.33 % |
| **Other sources of tinnitus variation*** | |
| Driving or when in a car | 24.32 % |
| Stress or anxiety | 67.57 % |
| When tired or very tired | 43.24 % |
| When I drink alcohol | 29.73 % |
| When I drink coffee | 2.7 % |
| When I take certain medications | 8.11 % |
| After exertion of sports | 21.62 % |
| Pressing on the head, neck or near the ear | 29.73 % |
| None of these situations | 13.51 % |

| Sleep parasomnia symptomatologic profile | Percentage |
|---|---|
| **Sleep parasomnia diagnoses*** | |
| Sleep apnea | 24.32 % |
| Narcolepsy | 0.0 % |
| Sleepwalking; night terror | 0.0 % |
| Hypersomnia | 0.0 % |
| Behavioral disorder during REM sleep | 2.7 % |
| Restless legs syndrome | 18.92 % |
| Bruxism (teeth grinding during sleep) | 62.16 % |
| Other | 8.11 % |
| **Effect of Sleeping Position on Tinnitus*** | |
| Increase, when on my back | 2.7 % |
| Increase, when on my left side | 5.41 % |
| Increase, when on the right side | 8.11 % |
| Increase, when on the left side | 2.7 % |
| Increase, when I sleep in a seated position | 5.41 % |
| Decrease, when on my stomach | 2.7 % |
| Decrease, when on my back | 10.81 % |
| Decrease, when on the left side | 2.7 % |
| Decrease, when on the right side | 2.7 % |
| Decrease, when on my stomach | 2.7 % |
| Decrease, when I sleep in a seated position | 2.7 % |
| No influence of position on my tinnitus | 83.78 % |
| **Average hours of sleep per night** | |
| Less than 6 hours per night | 16.22 % |
| Between 6 to 8 hours | 72.97 % |
| More than 8 hours per night | 10.81 % |
| **Difficulty falling asleep or staying asleep** | |
| No | 45.95 % |
| Yes, in staying asleep | 40.54 % |
| Yes, in falling asleep | 5.41 % |
| Yes, both | 8.11 % |
| **Snoring** | |
| Yes | 51.35 % |
| No | 29.73 % |
| Don't know | 18.91 % |

(*) For these questions, responders could select multiple answers among the proposed choices

*Table 2 – Sample symptomatologic characterization N=37, Abbreviations : VNS : Visual Analog Scale*

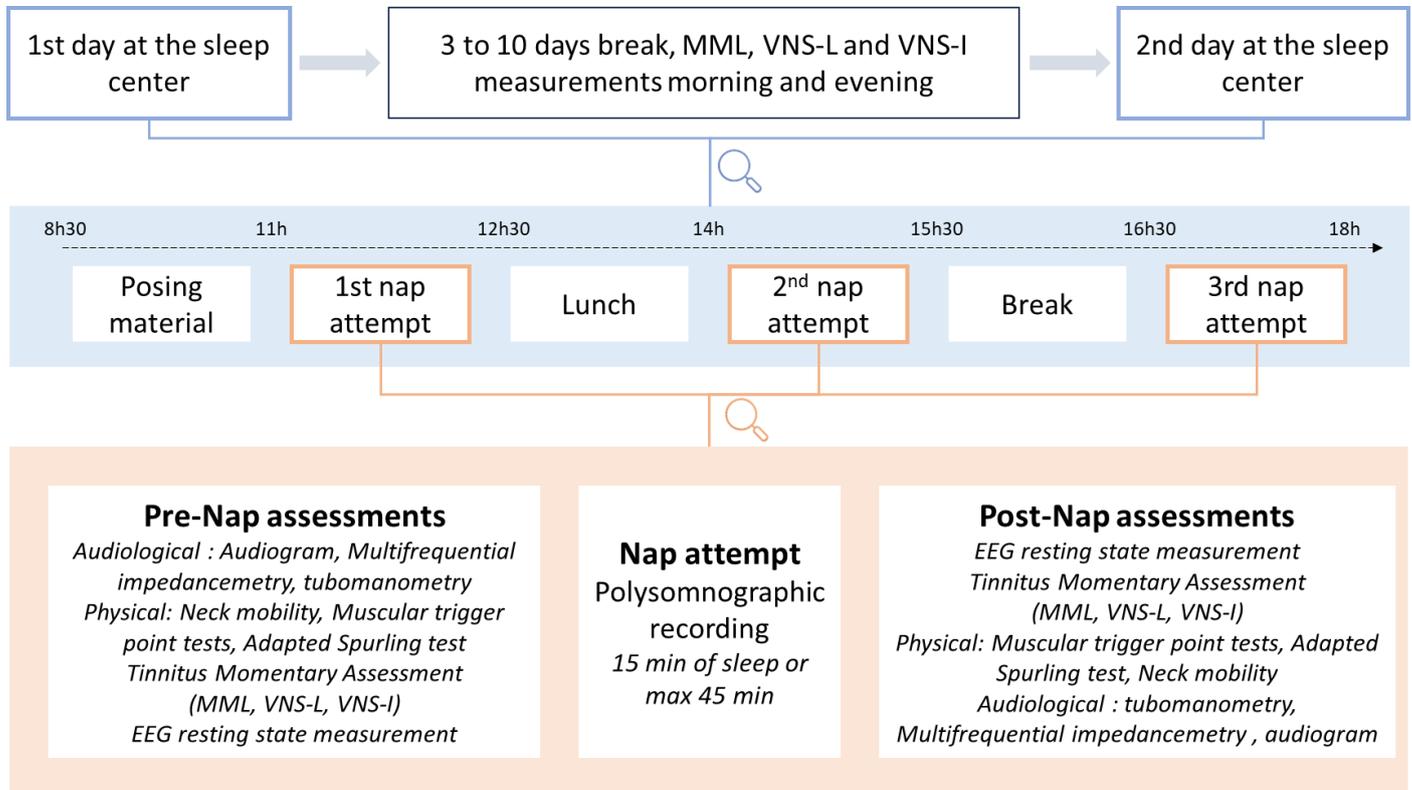

*Figure 1: Pipeline of participation of Tinninap subjects, MML : minimum masking level, VNS-L: visual numeric scale on tinnitus loudness, VNS-I: visual numeric scale on tinnitus intrusiveness, EEG: electroencephalography*

# Figure 2: Distribution of Z-scored Minimum Masking levels between before the nap and after the nap

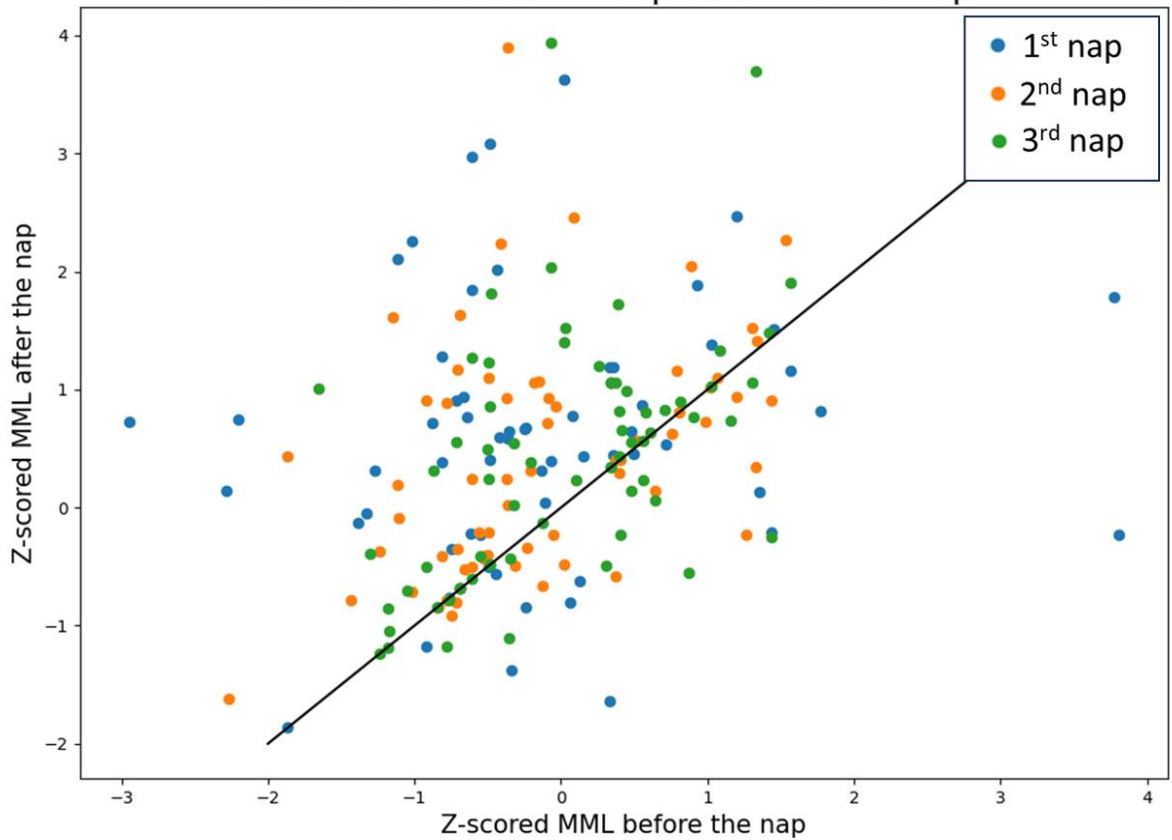

*Figure 2: Distribution of Z-scored Minimum Masking levels between before the nap and after the nap. Z-scoring of the minimum masking level was performed on all MML measurements before and after the six naps attempts, as well as the measurements performed at home each morning and each evening between the two days of participation, as described in the Tinnitus Momentary Assessment (TMA) subsection of the clinical assessment section (2.3) of the methods. Blue dots represent the distribution of the 1st nap, orange the 2nd nap and green the 3rd nap.*

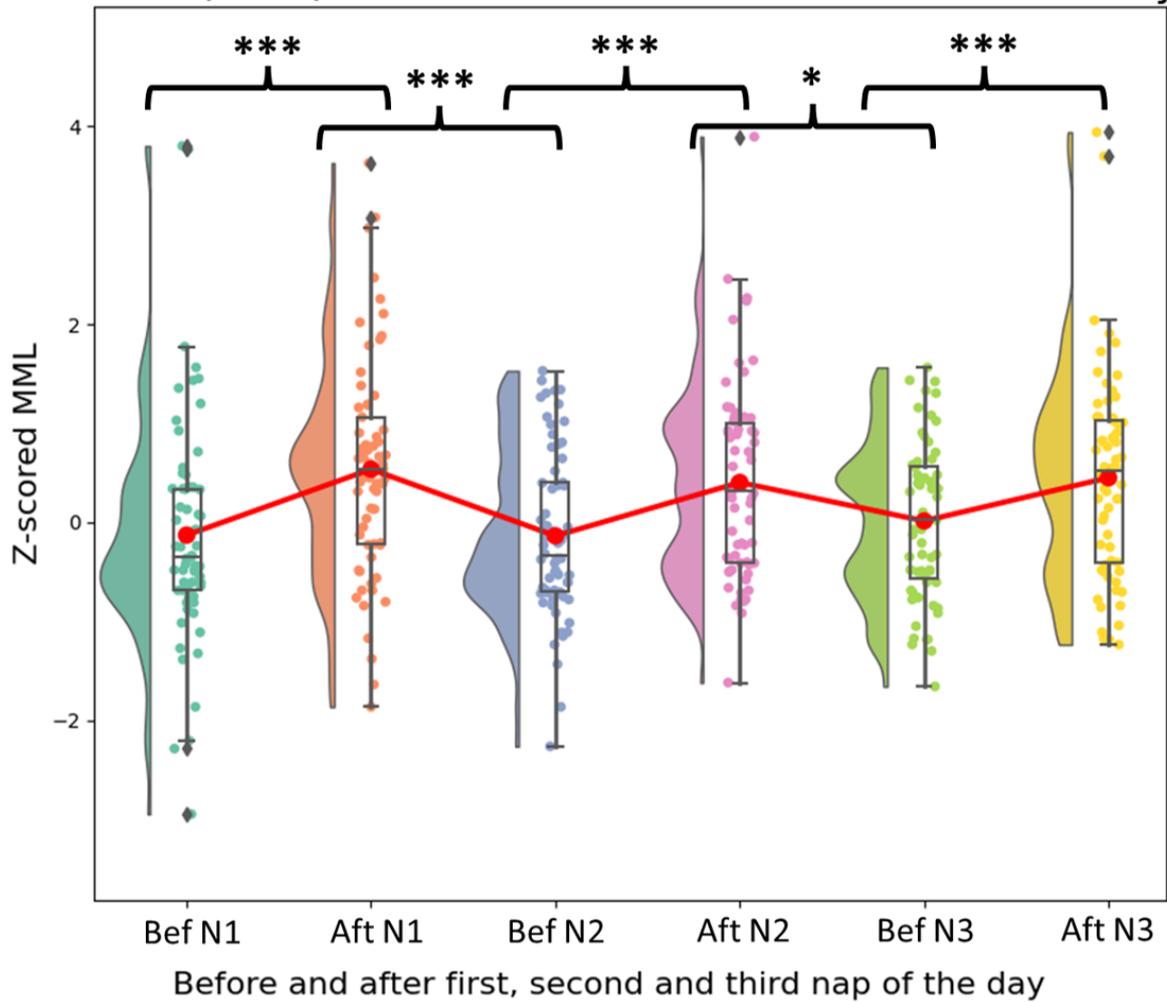

*Figure 3: Evolution of Z-scored Minimum Masking Level (MML) across time of measurement in the day. Bef: Before, Aft : After N1 : nap 1, N2 : nap 2, N3 : nap 3*

# Figure 4: Evolution of absolute differences of Z-scored Minimum Masking Level (MML) across time of measurement in the day

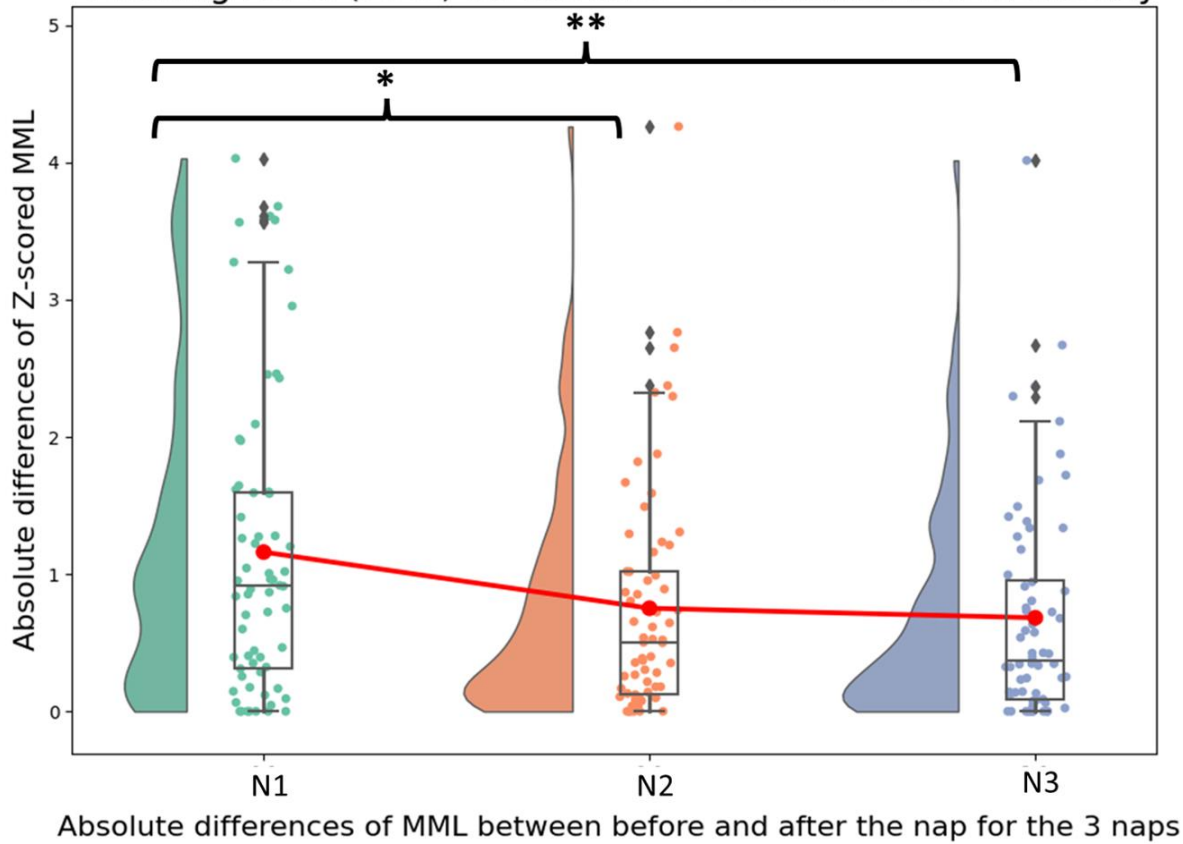

Absolute differences of MML between before and after the nap for the 3 naps

*Figure 4: Evolution of absolute differences of Z-scored Minimum Masking Level (MML) across time of measurement in the day. N1 : nap 1, N2 : nap 2, N3 : nap 3*

Table 3, Polysomnographic correlates of nap-induced tinnitus modulations

| Polysomnographic correlates | Number of patients | Delta MML | | Delta VNS-L | | Delta VNS-I | |
|---|---|---|---|---|---|---|---|
| | | Correlation dir | P-value | Correlation dir | P-value | Correlation dir | P-value |
| N2 sleep duration | 29 | Positive | 1.0 | Positive | **0.011*** | Positive | 1.0 |
| N2 and N3 sleep duration | 29 | Positive | 1.0 | Positive | **0.02*** | Positive | 1.0 |
| Snoring duration | 29 | Positive | **p < 0.001 *** | Positive | 0.38 | Positive | 0.75 |
| Average snore db | 29 | Positive | 0.2 | Positive | 0.12 | Positive | **0.012 *** |
| Apnea count | 29 | Negative | 1.0 | Positive | 0.14 | Positive | **p < 0.001 *** |
| SPO2 desaturation count | 25 | Negative | 0.56 | Positive | 0.75 | Positive | 0.32 |
| Supine position duration | 28 | Negative | 1.0 | Positive | 1.0 | Negative | 1.0 |
| Ipsilateral rotated sleep position | 28 | Positive | 1.0 | Negative | 0.26 | Positive | 1.0 |
| Controlateral rotated sleep position | 28 | Positive | 1.0 | Negative | 1.0 | Negative | 1.0 |
| Bruxism burst events count | 28 | Positive | 1.0 | Positive | 0.17 | Negative | 0.79 |
| *Abbreviations : VNS : Visual Numeric Scale, Correlation dir : direction of the correlation, (\*) : p < 0.05, (\*\*) : p < 0.01, (\*\*\*) : p < 0.001* | | | | | | | |

*Table 3 : Polysomnographic correlates of nap-induced tinnitus modulations. MML : Minimum Masking Level, VNS-L : Visual Analog Scale on tinnitus Loudness, VNS-I : Visual Analog Scale on tinnitus Intrusiveness, SPO2 : Blood-oxygen saturation level, dir : direction (of the correlation)*

Table 4, Audiologic and kinesiologic correlates of nap-induced tinnitus modulations (N =30)

| Audiologic and kinesiologic correlates | MML | | VNS-L | | VNS-I | |
|---|---|---|---|---|---|---|
| | Correlation dir | P-value | Correlation dir | P-value | Correlation dir | P-value |
| Cumulated muscular sensibility ipsi | Positive | 0.33 | Positive | 1.0 | Positive | 1.0 |
| Cumulated muscular sensibility contro | Positive | 0.12 | Positive | 1.0 | Positive | 1.0 |
| Cumulated tinnitus modulation ipsi | Negative | 1.0 | Positive | 1.0 | Positive | 1.0 |
| Cumulated tinnitus modulation contro | Negative | 1.0 | Positive | 1.0 | Positive | 1.0 |
| Neck extension-flexion amplitude (°) | Positive | 1.0 | Positive | 1.0 | Positive | 1.0 |
| Neck lateral flexion amplitude (°) | Negative | 1.0 | Positive | 1.0 | Positive | 0.64 |
| Neck rotation amplitude (°) | Negative | 1.0 | Negative | 1.0 | Positive | 1.0 |
| Jaw opening amplitude (cm) | Negative | 1.0 | Negative | 1.0 | Negative | 1.0 |
| Jaw diduction amplitude ipsi (cm) | Positive | 1.0 | Positive | 1.0 | Positive | 1.0 |
| Jaw diduction amplitude contro (cm) | Negative | 1.0 | Positive | 1.0 | Positive | 1.0 |
| Intensity threshold closest freq ipsi (dB) | Positive | 1.0 | Negative | 1.0 | Negative | 1.0 |
| Intensity threshold closest freq contro (dB) | Positive | 1.0 | Positive | 1.0 | Positive | 0.26 |
| Average tone loss | Positive | 1.0 | Positive | 1.0 | Positive | 0.62 |
| Multifreq Imp Resonance freq ipsi | Negative | 0.91 | Negative | 1.0 | Negative | 1.0 |
| Multifreq Imp Peak pressure ipsi | Positive | 1.0 | Positive | 1.0 | Positive | 1.0 |
| Multifreq Imp Resonance freq contro | Negative | 1.0 | Negative | 0.79 | Negative | 1.0 |
| Multifreq Imp Peak pressure contro | Positive | 1.0 | Positive | 1.0 | Positive | 1.0 |

*Abbreviations : Correlation dir : direction of the correlation, ipsi : measurement performed on the side of tinnitus perception, ipsi : measurement performed on the side controlateral to tinnitus perception, freq : frequency, multifreq imp : multifrequential impedancemetry*

*Table 4 : Audiologic and kinesiologic correlates of nap-induced tinnitus modulations (N=30). MML : Minimum Masking Level, VNS-L : Visual Analog Scale on tinnitus Loudness, VNS-I : Visual Analog Scale on tinnitus Intrusiveness, dir : direction (of the correlation), ipsi : measurement performed on the side of the tinnitus perception, contro : measurement performed on the side contralateral to the tinnitus perception, multifreq imp : multifrequency impedancemetry.*

# Supplementary Material:

Hereafter in Supplementary Tables 1 and 2 are presented the results of correlational tests between Tinnitus Momentary Assessment (TMA) features (Minimum masking level, Visual analog scales on tinnitus loudness and intrusiveness) and a more extended set of intra-nap (Table 1) and extra-nap (Table 2) measurements, as well as with the detailed sleep apnea scoring features independently scored by Mrs C.G. cited in the Acknowledgments section. Please note that these tests were purely exploratory, this is why no statistical corrections for multiple testing were applied here.

Hereafter are described the meaning and procedure of obtention of the extended features presented in these tables.

**Extended Intra-nap features**

Snore and apnea density: it was calculated as the sum of durations of snores, apneas and hypopneas events during the nap divided by the duration of sleep during the nap.

Head positions durations per nap were manually scored from video recordings of the nap sessions. They were calculated as percentages of the total duration of the nap, as for body positions.

Distinction between sleep bruxism burst events and episodes : applying scoring recommendations from Lavigne and al (Lavigne et al., 1996), bursts of bruxism are 250 ms bursts of EMG activity in the suprahyoid and/or masseter muscles, without concomitant body movement. Bruxism episodes were scored from bursts with three different types : phasic (rhythmic), tonic (sustained), or mixed (both phasic and tonic). A phasic episode corresponds to at least three EMG bursts of 0.25 to 2.0 seconds' duration, separated by two inter burst intervals. A tonic episode corresponds to an EMG burst lasting more than 2.0 s.

**Extended extra-nap features**

Extended extra-nap features comprise the details of trigger point sensitivity and somatosensory modulations tests for all specific facial and cervical muscles tested, for both ipsilateral and contralateral sides. It also includes results of the adapted Spurling test and whether or not it elicited somatosensory modulations.

Likewise, correlational tests between all intensity thresholds frequencies for both sides ipsilateral and contralateral to tinnitus are detailed.

Correlations tests between TMA and tinnitus furthest frequency auditory thresholds were provided as a control test for closest frequency correlation test. Furthest frequency auditory threshold was the frequency among the ones tested that maximized the distance with tinnitus main frequency calculated from the tinnitus matching procedure.

Two supplementary features are also provided for multifrequency impedancemetry: separation frequency and separation pressure. Normal profiles of multifrequency impedancemetry frequency slices have only one maximum in pressure. This is what is observed classically in tympanograms curves. Yet, in pressure curves of slides of increasing the frequency, gradually the unique maximum in pressure splits in two maxima of pressure. This generally happens between 750 Hz and 3000 Hz and was described in (Torrecilla and Avan, 2021). Separation frequency and pressure was calculated from a custom-made automatic algorithmic procedure as the coordinates in the pressure/frequency plan of the separation between one maximum to two maxima.

**Second independent sleep apnea scoring**

In Supplementary Table 3 are presented exploratory analyses on more precise sleep apnea features performed by a second and independent sleep apnea scorer C.G.. Investigated metrics included : total count, accumulated duration and average duration sleep central, obstructive apnea and hypopnea events, as well as all event type average event duration and accumulated total duration. Last, accumulated duration including all apnea events type and snoring was included.

**Analysis of the relation between tinnitus severity and NITM**

In Supplementary figure 1, we analyzed if there was a relation between reported grades of THI scores in the initial assessment (CRF) and the average amplitude of NITM measured by MML variation over the nap. As displayed, no clear tendency appears. As a consequence, in our sample it appears that NITM amplitude is not related to tinnitus severity.

**Analysis of the relation between MML and hearing loss assessment**

In Supplementary figures 2 and 3, we tried to evaluate the relation between the degree of hearing loss assessed by the audiogram and the reported MML. Figure 1 represents the average auditory thresholds closest to tinnitus frequency and ipsilateral to tinnitus in dB as a function of average MML in dB. We observed a significant Spearman correlation between these measures for our sample of 37 patients ($p < 0.01$). Figure 2 represents the average hearing loss in dB as a function of average MML in dB. We observed a significant Spearman correlation between these measures for our sample of 37 patients ($p < 0.05$).

**Analysis of the physiological effect of naps on extra-nap measurements**

In order to evaluate if naps had in themselves a physiological effect on the metrics of interest, an overall comparison between the values before and after the naps was carried independently of the effect on TMA features. As described in the 2.5 section "Statistical analyses": the effect of naps on extra-nap measurements was assessed using a Fisher combination of p-value method over the list of one-sided paired Wilcoxon tests p-values for each participant scores before and after each nap. The results of this analysis are reported in Supplementary table 4, without statistical corrections. "Higher" means the values of the designated measurements were tested to be superior after the nap than they were before the nap. "Lower" means the values of the designated measurements were tested to be inferior after the nap than they were before the nap.

Interestingly, we observe a tendency of a shift in the audiogram thresholds values : both ipsilaterally and contralaterally thresholds below 3000 Hz seemed to improve after the nap while thresholds over 3000 Hz tended to worsen, which was congruent to a tendency of a reduction in resonance frequency of multifrequency impedancemetry.

Supplementary Table 1, Extended polysomnographic correlates of nap-induced tinnitus modulations

| | Number of patients | Delta MML | | Delta VNS-L | | Delta VNS-I | |
|---|---|---|---|---|---|---|---|
| | | Correlation dir | Pvalue | Correlation dir | Pvalue | Correlation dir | Pvalue |
| N1 duration | 29 | Negative | 0.066 | Negative | 0.55 | Positive | 0.995 |
| N2 duration | 29 | Positive | 0.358 | **Positive** | **0.001 \*\*** | Positive | 0.249 |
| N2+N3 duration | 29 | Positive | 0.373 | **Positive** | **0.002 \*\*** | Positive | 0.278 |
| Nap duration | 29 | Negative | 0.801 | Positive | 0.449 | Positive | 0.315 |
| Snore duration | 29 | **Positive** | **p < 0.001 \*\*\*** | Positive | 0.095 | Positive | 0.107 |
| Average snore sound intensity during nap | 29 | **Positive** | **0.023 \*** | **Positive** | **0.015 \*** | **Positive** | **0.001 \*\*** |
| Average room sound intensity during nap | 29 | Positive | 0.598 | Positive | 0.783 | Negative | 0.417 |
| Apnea count | 29 | Negative | 0.703 | **Positive** | **0.024 \*** | **Positive** | **p < 0.001 \*\*\*** |
| Apnea duration | 29 | Negative | 0.51 | Positive | 0.186 | **Positive** | **0.001 \*\*** |
| Snore and Apnea duration | 29 | **Positive** | **0.004 \*\*** | **Positive** | **0.007 \*\*** | **Positive** | **0.001 \*\*** |
| Snore and apnea density | 25 | Positive | 0.068 | Positive | 0.6 | **Positive** | **p < 0.001 \*\*\*** |
| SPO2 count | 28 | Negative | 0.07 | Positive | 0.251 | **Positive** | **0.04 \*** |
| Head upright | 29 | Negative | 0.885 | **Negative** | **0.002 \*\*** | **Negative** | **0.006 \*\*** |
| Head bent forward | 29 | Positive | 0.811 | Positive | 0.371 | Positive | 0.252 |
| Head bent backward | 29 | Negative | 0.738 | Negative | 0.735 | Negative | 0.723 |
| Head rotated ipsilateral to tinnitus | 29 | Negative | 0.615 | Positive | 0.061 | Negative | 0.949 |
| Head rotated controlateral to tinnitus | 29 | Negative | 0.198 | Positive | 0.399 | Negative | 0.188 |
| Head tilted ipsilateral to tinnitus | 29 | Positive | 0.15 | **Positive** | **0.002 \*\*** | **Positive** | **0.047 \*** |
| Head tilted controlateral to tinnitus | 29 | Positive | 0.424 | **Positive** | **0.004 \*\*** | Positive | 0.452 |
| Ipsilateral rotated sleep position | 29 | Positive | 0.251 | Negative | 0.95 | Positive | 0.571 |
| Controlateral rotated sleep position | 29 | Positive | 0.329 | Negative | 0.058 | Negative | 0.463 |
| Position supine | 29 | Negative | 0.34 | Positive | 0.905 | Negative | 0.653 |
| Bruxism burst events count | 28 | Positive | 0.725 | **Positive** | **0.026 \*** | Negative | 0.131 |
| Bruxism episodes count | 28 | **Positive** | **0.028\*** | **Positive** | **0.002 \*\*** | Positive | 0.179 |
| Bruxism density of burst events | 25 | **Negative** | **0.022 \*** | Negative | 0.118 | **Negative** | **0.003 \*\*** |
| *Abbreviations : VNS : Visual Numeric Scale, Correlation dir : direction of the correlation, (\*) : p < 0.05, (\*\*) : p < 0.01, (\*\*\*) : p < 0.001* | | | | | | | |

*Supplementary Table 1 : Extended polysomnographic correlates of nap-induced tinnitus modulations. MML : Minimum Masking Level, VNS-L : Visual Analog Scale on tinnitus Loudness, VNS-I : Visual Analog Scale on tinnitus Intrusiveness, SPO2 : Blood-oxygen saturation level, dir : direction (of the correlation)*

Supplementary Table 2, Extended audiologic and kinesiologic correlates of nap-induced tinnitus modulations (N=30)

| Audiologic and kinesiologic correlates | MML Correlation dir | MML P-value | VNS-L Correlation dir | VNS-L P-value | VNS-I Correlation dir | VNS-I P-value |
|---|---|---|---|---|---|---|
| Neck extensionflexion amplitude (°) | Positive | 0.971 | Positive | 0.636 | Positive | 0.205 |
| Neck lateral flexion amplitude (°) | Negative | 0.625 | Positive | 0.073 | **Positive** | **0.043 *** |
| Neck rotation amplitude (°) | Negative | 0.129 | Negative | 0.413 | Positive | 0.741 |
| TP sensibility trapezoid ipsilateral | Positive | 0.119 | Negative | 0.43 | Negative | 0.65 |
| TP sensibility SCM ipsilateral | Positive | 0.078 | Positive | 0.278 | Positive | 0.578 |
| TP sensibility splenius ipsilateral | Positive | 0.189 | Positive | 0.321 | Positive | 0.162 |
| TP sensibility masseter ipsilateral | Positive | 0.245 | Negative | 0.969 | Positive | 0.993 |
| TP sensibility temporal ipsilateral | Positive | 0.09 | Positive | 0.477 | Positive | 0.385 |
| TP sensibility trapezoid controlateral | **Positive** | **0.008 *** | Negative | 0.482 | Negative | 0.825 |
| TP sensibility SCM controlateral | **Positive** | **0.03 *** | Positive | 0.358 | Positive | 0.723 |
| TP sensibility splenius controlateral | Positive | 0.425 | Positive | 0.352 | Positive | 0.164 |
| TP sensibility masseter controlateral | Positive | 0.109 | Positive | 0.81 | Positive | 0.73 |
| TP sensibility temporal controlateral | Positive | 0.187 | Positive | 0.437 | Positive | 0.232 |
| Jaw opening amplitude (cm) | Negative | 0.78 | Negative | 0.381 | Negative | 0.992 |
| Jaw diduction amplitude ipsi (cm) | Positive | 0.351 | Positive | 0.39 | Positive | 0.669 |
| Jaw diduction amplitude contro (cm) | Negative | 0.793 | Positive | 0.383 | Positive | 0.587 |
| Adapted spurling test ipsilateral | Positive | 0.388 | Positive | 0.112 | **Positive** | **0.048 *** |
| Adapted spurling test controlateral | Positive | 0.154 | Positive | 0.574 | Positive | 0.195 |
| TP SM trapezoid ipsilateral | Negative | 0.256 | Negative | 0.276 | Negative | 0.143 |
| TP SM SCM ipsilateral | Positive | 0.996 | Negative | 0.769 | Positive | 0.944 |
| TP SM splenius ipsilateral | Positive | 0.406 | Negative | 0.513 | Negative | 0.386 |
| TP SM masseter ipsilateral | Positive | 0.411 | Negative | 0.41 | Negative | 0.4 |
| TP SM temporal ipsilateral | Positive | 0.268 | Negative | 0.402 | Negative | 0.371 |
| TP SM trapezoid controlateral | Negative | 0.393 | Negative | 0.447 | Negative | 0.432 |
| TP SM SCM controlateral | Negative | 0.905 | Negative | 0.894 | Negative | 0.995 |
| TP SM splenius controlateral | Positive | 0.428 | Positive | 0.545 | Negative | 0.356 |
| TP SM masseter controlateral | Positive | 0.769 | Negative | 0.369 | Negative | 0.343 |
| TP SM temporal controlateral | Positive | 0.395 | Negative | 0.534 | Positive | 0.644 |
| Jaw opening SM | Positive | 0.094 | Positive | 0.438 | Positive | 0.515 |
| Jaw diduction SM ipsi (cm) | Positive | 0.775 | Negative | 0.862 | Positive | 0.713 |
| Jaw diduction SM contro (cm) | Positive | 0.235 | Positive | 0.965 | Positive | 0.803 |
| Adapted spurling test SM ipsilateral | Positive | 0.757 | Negative | 0.491 | Negative | 0.215 |
| Adapted spurling test SM controlateral | Positive | 0.931 | Negative | 0.503 | Negative | 0.123 |
| Intensity threshold closest freq ipsi (dB) | Positive | 0.573 | Negative | 0.619 | Negative | 0.899 |
| Intensity threshold closest freq contro (dB) | Positive | 0.133 | Positive | 0.155 | **Positive** | **0.015 *** |
| Intensity threshold furthest freq ipsi (dB) | Positive | 0.971 | Positive | 0.921 | Negative | 0.777 |
| Intensity threshold furthest freq contro (dB) | Positive | 0.379 | Positive | 0.948 | Negative | 0.782 |
| Intensity threshold 125 Hz freq ipsi (dB) | Positive | 0.161 | Positive | 0.501 | Positive | 0.649 |
| Intensity threshold 250 Hz freq ipsi (dB) | Negative | 0.939 | Negative | 0.586 | Positive | 0.758 |
| Intensity threshold 500 Hz freq ipsi (dB) | Positive | 0.26 | Positive | 0.231 | Positive | 0.126 |
| Intensity threshold 750 Hz freq ipsi (dB) | Positive | 0.582 | Positive | 0.452 | Positive | 0.33 |
| Intensity threshold 1000 Hz freq ipsi (dB) | Positive | 0.521 | Positive | 0.174 | Positive | 0.154 |
| Intensity threshold 1500 Hz freq ipsi (dB) | Positive | 0.611 | Positive | 0.585 | Positive | 0.18 |
| Intensity threshold 2000 Hz freq ipsi (dB) | Positive | 0.525 | Negative | 0.921 | Positive | 0.898 |
| Intensity threshold 3000 Hz freq ipsi (dB) | Positive | 0.717 | Positive | 0.392 | Positive | 0.369 |
| Intensity threshold 4000 Hz freq ipsi (dB) | Positive | 0.557 | **Positive** | **0.043 *** | **Positive** | **0.009 *** |
| Intensity threshold 6000 Hz freq ipsi (dB) | Negative | 0.684 | Positive | 0.336 | Positive | 0.359 |
| Intensity threshold 8000 Hz freq ipsi (dB) | Positive | 0.933 | Negative | 0.724 | Positive | 0.961 |
| Intensity threshold 125 Hz freq contro (dB) | Positive | 0.097 | Positive | 0.108 | Positive | 0.361 |
| Intensity threshold 250 Hz freq contro (dB) | Positive | 0.178 | Positive | 0.571 | Positive | 0.373 |
| Intensity threshold 500 Hz freq contro (dB) | Positive | 0.271 | Positive | 0.266 | Positive | 0.425 |
| Intensity threshold 750 Hz freq contro (dB) | Positive | 0.772 | Positive | 0.536 | Positive | 0.631 |
| Intensity threshold 1000 Hz freq contro (dB) | Negative | 0.973 | Positive | 0.218 | Positive | 0.515 |
| Intensity threshold 1500 Hz freq contro (dB) | Positive | 0.522 | Positive | 0.686 | Positive | 0.56 |
| Intensity threshold 2000 Hz freq contro (dB) | Positive | 0.728 | Negative | 0.753 | Positive | 0.792 |
| Intensity threshold 3000 Hz freq contro (dB) | Positive | 0.126 | Positive | 0.795 | Negative | 0.938 |
| Intensity threshold 4000 Hz freq contro (dB) | Positive | 0.538 | Positive | 0.791 | Positive | 0.619 |
| Intensity threshold 6000 Hz freq contro (dB) | Positive | 0.436 | Positive | 0.948 | Negative | 0.954 |
| Intensity threshold 8000 Hz freq contro (dB) | Positive | 0.412 | Positive | 0.545 | Positive | 0.063 |
| Average tone loss | Positive | 0.235 | Positive | 0.159 | **Positive** | **0.039 *** |
| Multifreq Imp Resonance freq ipsi (Hz) | Negative | 0.061 | Negative | 0.079 | Negative | 0.217 |
| Multifreq Imp Peak pressure ipsi (Pa) | Positive | 0.376 | Positive | 0.557 | Positive | 0.752 |
| Multifreq Imp Separation freq ipsi (Hz) | Negative | 0.556 | Positive | 0.325 | Positive | 0.194 |
| Multifreq Imp Separation pressure ipsi (Pa) | Positive | 0.473 | Positive | 0.357 | Positive | 0.889 |
| Multifreq Imp Resonance freq contro (Hz) | Negative | 0.803 | **Negative** | **0.046 *** | Negative | 0.089 |
| Multifreq Imp Peak pressure contro (Pa) | Positive | 0.709 | Positive | 0.481 | Positive | 0.447 |
| Multifreq Imp Separation freq contro (Hz) | Positive | 0.664 | Negative | 0.581 | Negative | 0.182 |
| Multifreq Imp Separation pressure contro (Pa) | Positive | 0.859 | Positive | 0.782 | Positive | 0.785 |

*Abbreviations : Correlation dir : direction of the correlation, ipsi : measurement performed on the side of tinnitus perception, contro : measurement performed on the side controlateral to tinnitus perception, freq : frequency, multifreq imp : multifrequency impedancemetry, TP : trigger point, SCM : sternocleidomastoid, SM : somatosensory modulation*

*Supplementary Table 2 : Extended audiologic and kinesiologic correlates of nap-induced tinnitus modulations (N=30). MML : Minimum Masking Level, VNS-L : Visual Analog Scale on tinnitus Loudness, VNS-I : Visual Analog Scale on tinnitus Intrusiveness, dir : direction (of the correlation), ipsi : measurement performed on the side of the tinnitus perception, contro : measurement performed on the side contralateral to the tinnitus perception, multifreq imp : multifrequency impedancemetry, TP :trigger point, SCM : sternocleidomastoid, SM : somatosensory modulation*

Supplementary Table 3, Extended sleep apnea scoring correlates of nap-induced tinnitus modulations from scorer 2

| | Number of patients | Delta MML | | Delta VNS-L | | Delta VNS-I | |
|---|---|---|---|---|---|---|---|
| | | Correlation dir | Pvalue | Correlation dir | Pvalue | Correlation dir | Pvalue |
| Central sleep apnea count | 29 | Positive | 0.917 | Negative | 0.824 | Positive | 0.851 |
| Central sleep apnea duration | 29 | Positive | 0.731 | Positive | 0.888 | Positive | 0.679 |
| Central sleep apnea average event duration | 29 | Negative | 0.941 | Positive | 0.723 | Positive | 0.822 |
| Obstructive sleep apnea count | 29 | Negative | 0.662 | Negative | 0.302 | Positive | 0.111 |
| Obstructive sleep apnea duration | 29 | Negative | 0.81 | Negative | 0.181 | Positive | 0.275 |
| Obstructive sleep apnea average event duration | 29 | Negative | 0.454 | Negative | 0.167 | Positive | 0.682 |
| Sleep hypopnea count | 30 | Positive | 0.203 | **Positive** | **0.001\*\*** | **Negative** | **0.001\*\*** |
| Sleep hypopnea duration | 30 | Positive | 0.162 | **Positive** | **0.001\*\*** | **Positive** | **0.001\*\*** |
| Sleep hypopnea average event duration | 30 | Negative | 0.308 | Positive | 0.579 | Negative | 0.255 |
| All event type average event duration | 30 | Negative | 0.417 | Negative | 0.675 | Negative | 0.682 |
| All event total duration | 30 | Positive | 0.878 | Positive | 0.104 | Positive | 0.079 |
| Snore and apnea duration | 30 | **Positive** | **0.001\*\*** | Positive | 0.104 | **Positive** | **0.001\*\*** |
| *Abbreviations : VNS : Visual Numeric Scale, Correlation dir : direction of the correlation, (\*) : p < 0.05, (\*\*) : p < 0.01, (\*\*\*) : p < 0.001* | | | | | | | |

*Supplementary Table 3 : Extended sleep apnea scoring correlates from scorer 2. MML : Minimum Masking Level, VNS-L : Visual Analog Scale on tinnitus Loudness, VNS-I : Visual Analog Scale on tinnitus Intrusiveness, dir : direction (of the correlation).*

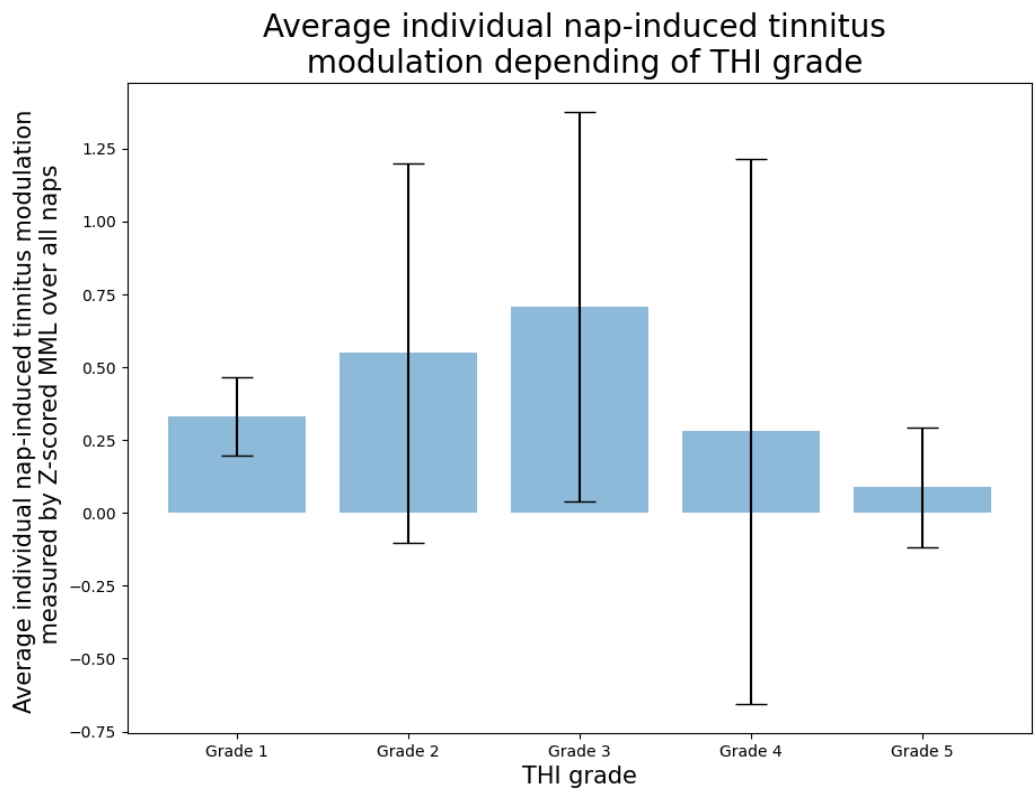

*Supplementary figure 1 – Amplitude of nap-induced tinnitus modulation, measured by the individual average Z-scored minimum masking level difference over all naps as a function of tinnitus severity, measured by THI grade. THI grades are defined as 0-16: Slight or no handicap (Grade 1), 18-36: Mild handicap (Grade 2), 38-56: Moderate handicap (Grade 3), 58-76: Severe handicap (Grade 4), 78-100: Catastrophic handicap (Grade 5)*

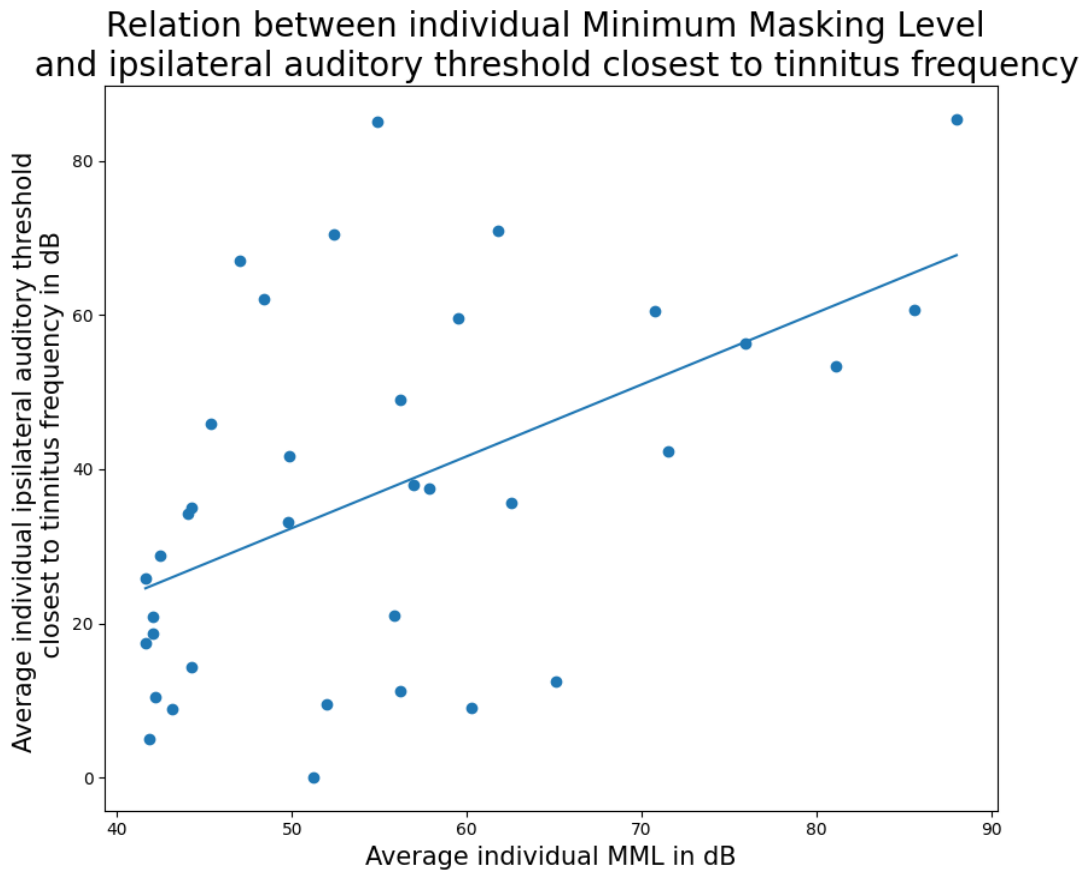

*Supplementary figure 2 – in this representation, for each patient we displayed the average auditory threshold in dB closest to tinnitus frequency and ipsilateral to tinnitus as a function of average MML in dB. MML : Minimum Masking Level*

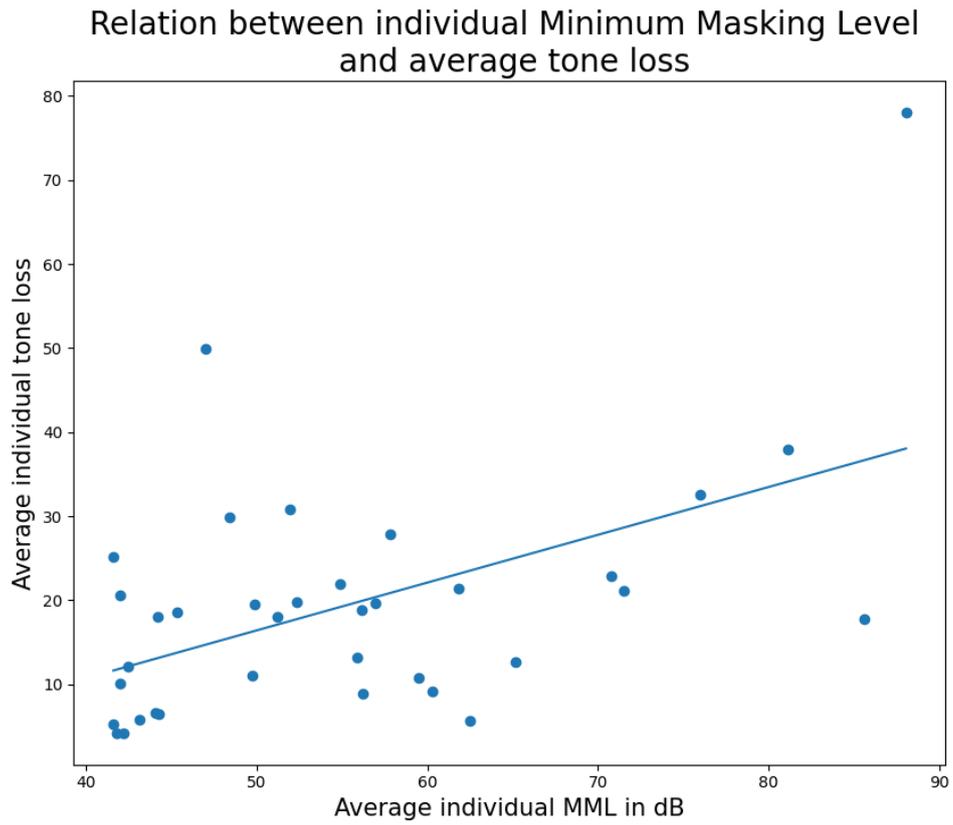

*Supplementary figure 3 – in this representation, for each patient we displayed the average tone loss in dB as a function of average MML in dB. MML : Minimum Masking Level*

## Supplementary table 4, Physiological effect of naps on extra-nap measurements

| Extra-nap measurement name | P-value test higher | P-value test lower | Extra-nap measurement name | P-value test higher | P-value test lower |
|---|---|---|---|---|---|
| Neck extensionflexion amplitude (°) | 0,993 | 0,539 | Jaw opening amplitude (cm) | 0,114 | 0,963 |
| Neck lateral flexion amplitude (°) | 0,994 | 0,43 | Jaw opening SM | 0,093 | 0,888 |
| Neck rotation amplitude (°) | 0,684 | 0,933 | Average tone loss | 0,914 | 0,077 |
| TP sensibility trapezoid ipsilateral | 0,96 | **0,026 *** | TP sensibility trapezoid controlateral | 0,777 | 0,184 |
| TP sensibility SCM ipsilateral | 0,214 | 0,291 | TP sensibility SCM controlateral | **0,046 *** | 0,94 |
| TP sensibility splenius ipsilateral | 0,988 | **0,01 *** | TP sensibility splenius controlateral | 0,991 | **0,028 *** |
| TP sensibility masseter ipsilateral | 0,872 | 0,426 | TP sensibility masseter controlateral | 0,92 | 0,462 |
| TP sensibility temporal ipsilateral | 0,999 | **0,023 *** | TP sensibility temporal controlateral | 0,973 | 0,134 |
| Jaw diduction amplitude ipsi (cm) | 0,287 | 0,547 | Jaw diduction amplitude contro (cm) | 0,381 | 0,415 |
| Adapted spurling test ipsilateral | 0,9 | 0,229 | Adapted spurling test controlateral | 0,74 | 0,253 |
| TP SM trapezoid ipsilateral | 0,225 | 0,675 | TP SM trapezoid controlateral | 0,237 | 0,505 |
| TP SM SCM ipsilateral | 0,49 | 0,49 | TP SM SCM controlateral | 0,666 | 0,266 |
| TP SM splenius ipsilateral | 0,065 | 0,862 | TP SM splenius controlateral | 0,311 | 0,518 |
| TP SM masseter ipsilateral | 0,143 | 0,903 | TP SM masseter controlateral | **0,013 *** | 1 |
| TP SM temporal ipsilateral | 0,354 | 0,627 | TP SM temporal controlateral | 0,089 | 0,937 |
| Jaw diduction SM ipsi (cm) | 0,547 | 0,453 | Jaw diduction SM contro (cm) | 0,901 | 0,091 |
| Adapted spurling test SM ipsilateral | 0,421 | 0,852 | Adapted spurling test SM controlateral | 0,637 | 0,522 |
| Intensity threshold closest freq ipsi (dB) | 0,083 | 0,971 | Intensity threshold closest freq contro (dB) | 0,299 | 0,747 |
| Intensity threshold furthest freq ipsi (dB) | 0,85 | **0,042 *** | Intensity threshold furthest freq contro (dB) | 0,76 | **0,036 *** |
| Intensity threshold 125 Hz freq ipsi (dB) | 0,765 | 0,064 | Intensity threshold 125 Hz freq contro (dB) | 0,913 | 0,052 |
| Intensity threshold 250 Hz freq ipsi (dB) | 0,787 | 0,297 | Intensity threshold 250 Hz freq contro (dB) | 0,968 | **0,035 *** |
| Intensity threshold 500 Hz freq ipsi (dB) | 0,884 | 0,107 | Intensity threshold 500 Hz freq contro (dB) | 0,834 | 0,061 |
| Intensity threshold 750 Hz freq ipsi (dB) | 0,574 | 0,31 | Intensity threshold 750 Hz freq contro (dB) | 0,786 | 0,115 |
| Intensity threshold 1000 Hz freq ipsi (dB) | 0,57 | 0,447 | Intensity threshold 1000 Hz freq contro (dB) | 0,323 | 0,289 |
| Intensity threshold 1500 Hz freq ipsi (dB) | 0,49 | 0,368 | Intensity threshold 1500 Hz freq contro (dB) | 0,462 | 0,243 |
| Intensity threshold 2000 Hz freq ipsi (dB) | 0,765 | 0,251 | Intensity threshold 2000 Hz freq contro (dB) | 0,434 | 0,546 |
| Intensity threshold 3000 Hz freq ipsi (dB) | 0,85 | 0,057 | Intensity threshold 3000 Hz freq contro (dB) | 0,655 | 0,118 |
| Intensity threshold 4000 Hz freq ipsi (dB) | 0,146 | 0,941 | Intensity threshold 4000 Hz freq contro (dB) | 0,394 | 0,301 |
| Intensity threshold 6000 Hz freq ipsi (dB) | 0,161 | 0,656 | Intensity threshold 6000 Hz freq contro (dB) | 0,257 | 0,36 |
| Intensity threshold 8000 Hz freq ipsi (dB) | 0,07 | 0,955 | Intensity threshold 8000 Hz freq contro (dB) | 0,051 | 0,936 |
| Multifreq Imp Resonance freq ipsi (Hz) | 0,959 | 0,119 | Multifreq Imp Resonance freq contro (Hz) | 0,949 | 0,136 |
| Multifreq Imp Peak pressure ipsi (Pa) | 0,574 | 0,922 | Multifreq Imp Peak pressure contro (Pa) | 0,736 | 0,877 |
| Multifreq Imp Separation freq ipsi (Hz) | 0,997 | 0,183 | Multifreq Imp Separation freq contro (Hz) | 0,985 | 0,059 |
| Multifreq Imp Separation pressure ipsi (Pa) | 0,227 | 0,111 | Multifreq Imp Separation pressure contro (Pa) | 0,608 | 0,786 |

*Abbreviations : ipsi : measurement performed on the side of tinnitus perception, contro : measurement performed on the side controlateral to tinnitus perception, freq : frequency, multifreq : multifrequency impedancemetry, TP : trigger point, SCM : sternocleidomastoid, SM : somatosensory modulation*

*Supplementary Table 4 : Physiological effect of naps on extra-nap measurements. Fisher combination of one-sided paired Wilcoxon test p-values are reported in columns "higher" and "lower". "Higher" means the values of the designated measurements were tested to be superior after the nap than they were before the nap. "Lower" means the values of the designated measurements were tested to be inferior after the nap than they were before the nap.*

*Abbreviations : ipsi : measurement performed on the side of tinnitus perception, contro : measurement performed on the side contralateral to tinnitus perception, freq : frequency, multifreq imp : multifrequency impedancemetry, TP : trigger point, SCM : sternocleidomastoid, SM : somatosensory modulation*